\documentclass[12pt]{article}
\usepackage{amsmath,amssymb,amsfonts,epsfig,graphicx}
\newcommand{\be}{\begin{equation}}
\newcommand{\ee}{\end{equation}}

\newcommand{\bse}{\begin{subequations}}
\newcommand{\ese}{\end{subequations}}
\newcommand{\bea}{\begin{eqnarray}}
\newcommand{\eea}{\end{eqnarray}}
\newcommand{\ba}{\begin{array}}
\newcommand{\ea}{\end{array}}

\def\half{\frac{1}{2}}
\def\cN{{\cal N}}
\def\Ne{${\cal N}=8$\ }\def\Ns{${\cal N}=6$\ }
\def\3dNe{$3d,\ {\cal N}=8\ $}

\def\sun2{$su(N)\times su(N)$}
\def\cs{Chern-Simons }
\def\susy{supersymmetry}
\def\ta{three-algebra}
\def\FI{fundamental identity}

\def\bbl{\bigl{[}\hspace{-1.2mm}\bigl{[}}
\def\bbr{\bigr{]}\hspace{-1.1mm}\bigr{]}}
\def\id{1\!\!1}
 \makeatletter \@addtoreset{equation}{section}

\makeatother
\begin{document}
\baselineskip 18pt%

\begin{titlepage}
\vspace*{1mm}%
\hfill%
\vbox{
    \halign{#\hfil \cr
\;\;\;\;\;\;\;\;\;\; IPM/P-2008/049 \cr arXiv:0810.3782\tt{
[hep-th]} \cr
           \cr
           } 
      }  
\vspace*{10mm}%

\centerline{{\Large {\bf  \textsl{
A New Three-Algebra Representation for the}}}}
\centerline{{\Large {\bf \textsl{${\cal N}=6$ $su(N)\times su(N)$
Superconformal Chern-Simons Theory}}}}\vspace*{3mm}

\vspace*{7mm}
\begin{center}
{\bf \large{ M. M. Sheikh-Jabbari  }}
\end{center}
\begin{center}
\vspace*{0.4cm} {\it {School of Physics, Institute for research in
fundamental
sciences (IPM)\\
P.O.Box 19395-5531, Tehran, IRAN}}\\
{E-mail: {\tt jabbari @theory.ipm.ac.ir}}%
\vspace*{1.5cm}
\end{center}

\begin{center}{\bf Abstract}\end{center}
\begin{quote}


Based on the realization of three-algebras in terms of algebra of
matrices and four-brackets [arXiv:0807.1570] we present the notion
of $u(N)$-based extended  three-algebras, which for $N=2$ reproduces
 the Bagger-Lambert three-algebra. Using these extended
three-algebras we construct an $su(N)\times su(N)$ Chern-Simons
action with explicit $SO(8)$ invariance. The dynamical fields of
this theory are eight \emph{complex valued} bosonic and fermionic
fields in the bi-fundamental representation of the $su(N)\times
su(N)$. For generic $N$ the fermionic transformations, however,
close only on a subclass of the states of this theory onto the $3d$,
${\cal N}=6$ superalgebra. In this sector we deal with four complex
valued scalars and fermions, our theory is closely related to the
ABJM model [arXiv:0806.1218], and hence it can be viewed as the (low
energy effective) theory of $N$ M2-branes. We discuss that our
three-algebra structure suggests a picture of open M2-brane
stretched between any two pairs of M2-branes. We also analyze the
BPS configurations of our model.

\end{quote}
\end{titlepage}
\tableofcontents
\section{Introduction}

Motivated by the proposal made by J. Schwarz \cite{Schwarz-04},
recently Bagger and Lambert \cite{BL1, BL2} and Gustavson
\cite{Gust1, Gust2} have proposed an action for maximally
supersymmetric three-dimensional conformal field theory (see
\cite{Lambert-strings08} for a recent review). This action is
basically a supersymmetric Chern-Simons theory, in which instead
of the usual Lie-algebraic structures and commutators one deals
with a new type of algebra which has a bracket involving three
elements of the algebra (rather than two for the commutator). This
kind of algebra was hence called \emph{three-algebra}.

The metric \ta s are defined through a three-bracket structure and
a ``trace'' over the algebra (and hence a metric) and a
generalization of the Jacobi Identity, the \emph{fundamental
identity}. According to the \emph{three-algebra no-go theorem}
\cite{Papadopoulos:2008sk} the only three-algebra which has a
positive definite norm is either $so(4)$ or direct sums of a
number of $so(4)$'s. In this sense the original
Bagger-Lambert-Gustavson (BLG) theory is rather unique
\cite{Lambert-strings08}.

The restriction  were bypassed relaxing the positive norm
condition and it was shown \cite{Russo, Verlinde, HIM} (see also
\cite{Jose-Fig}) that allowing a single negative eigenvalue in the
metric one has the possibility of constructing \ta s based on
 any  Lie-algebra. The BLG theory based on these
\emph{Lorentzian} \ta s, due to the negative norm in the metric
has pathologic ghost-type fields (fields with negative kinetic
energy). Despite of the proposals and arguments that these
ghost-type fields are not harmful to the unitarity of the theory
\cite{Sen1, Sen2, ghost-gauging, ASS, VerlindeM2scattering} the
connection of these theories to that of multi M2-brane is not
clear yet.

The $3d,$ \Ne Super-Conformal Field Theory (SCFT) is expected to
arise from the low energy limit of a system of multi M2-branes and
be dual to M-theory on $AdS_4\times S^7$ \cite{AdS/CFT}. With this
motivation and the difficulties with extending the BLG theory and
their usual \ta s, inspired by ideas in \cite{Mark} \footnote{In
\cite{Mark} it was shown that the $so(4)$-based BLG theory is
nothing but an  $su(2)\times su(2)$ Chern-Simons theory with ${\cal
N}=8$ supersymmetry.}, Aharony, Bergman, Jafferis and Maldacena,
(ABJM) \cite{ABJM} constructed an ${\cal{N}}=6$ $u(N)\times u(N)$
supersymmetric Chern-Simons theory at level $k$ with matter fields
in the bi-fundamental of the gauge group. This theory is proposed to
be describing $N$ M2-branes on a $Z_k$ orbifold or M-theory on
$AdS_4\times S^7/Z_k$. In \cite{BL3-bracket-N6} it was shown that
the ABJM theory has a representation in terms of the BLG theory with
a ``generalized'' notion of \ta .

In this paper we attempt in writing an explicit action for the
$3d$, \Ne $su(N)\times su(N)$ Chern-Simons theory.  To this end we
start from the BLG theory but with a new \emph{extended \ta }.
Using the four-bracket representation for the \ta s introduced in
\cite{ASS} (see also \cite{TGMT}) we give a matrix representation
for the extended three-algebra in terms of $2N\times 2N$ Hermitian
matrices. The underlying $su(2N)$ algebra has an \sun2\
subalgebra. Utilizing this matrix representation we show that the
BLG theory  with the above ``$u(N)$-based extended \ta '' is
equivalent to a $3d$ \sun2 \ \cs action. We show that for the
$N=2$ case our extended three-algebra reproduces  two copies of
the Bagger-Lambert three-algebra. In this action, which for
generic $N$ has explicit global $SO(8)$ invariance, we are forced
to work with eight \emph{complex valued} scalars and fermions in
the bi-fundamental representation of the \sun2.

The direct generalization of 16 fermionic transformations of the BLG
theory, however, do not close onto a generic configuration of the
fields in our theory and hence our theory, despite of being $SO(8)$
invariant, is not an \Ne theory. One may then ask if there is a
subclass or a sector of physical configuration over which all or a
subset of fermionic transformations indeed form a \susy\ algebra. As
we will show for generic $N$  the largest of such sectors in the
Fock space of the theory is the part which is invariant under
$SU(4)\times U(1)\in SO(8)$, and with fermionic transformation
parameters restricted to be in $\mathbf{6}_0$ of this $SU(4)\times
U(1)$. In this sector the bosonic scalar degrees of freedom of the
theory are four complex valued fields in $\mathbf{4}_{+1}$ of
$SU(4)\times U(1)$ in bi-fundamental of \sun2 and their complex
conjugates, half of our original theory. In this sector the theory
exhibits \Ns \susy\ which is the largest possible \susy\ within the
class of our models and is hence closely related to the ABJM model
\cite{ABJM}. We show that for the special case of $N=2$, because of
the special properties of the $su(2)$ algebra, besides the
projection onto the $SU(4)\times U(1)$ sector, one has the option of
closing all 16 \susy\ variations by projecting into another
invariant sector while keeping the $SO(8)$. In this sense the
Bagger-Lambert theory is different than the ABJM theory for $N=2$.

We propose that our  \sun2 \cs theory once projected onto the
$SU(4)\times U(1)$ sector, describes the low energy theory for $N$
M2-branes on the flat space background. Our construction in terms of
$N\times N$ complex valued fields finds a natural suggestive
``geometric'' picture through \emph{two} pairs of open membranes
stretched between any two M2-branes. These two pairs are related by
the $3d$ worldvolume parity which is connected with the
``projection'' onto the $SU(4)\times U(1)$ invariant sector in the
Fock space described above. This picture sheds light on both the
underlying $2N\times 2N$ matrices and $su(2N)$ structure, its \sun2
subalgebra and why the projection is necessary to avoid over
counting of degrees of freedom.

The paper is organized as follows. In section 2, we review basics
of three-algebras and their representation in terms of ordinary
matrices and the four-brackets. In section 3, we  present the
notion of ``extended'' three-algebra and also the
\emph{$u(N)$-based extended three-algebra}, the three-algebra that
we propose for $N$ M2-bane theory. In section 4, we construct the
BLG theory based on the extended three-algebra and discuss its
supersymmetry, gauge symmetry and other global symmetries as well
as the behavior under the $3d$ parity. In section 5, we show that
our theory is equivalent to an \sun2 \cs gauge theory with
explicit $SO(8)$ invariance, while not \Ne invariant. We discuss
its relation to the ABJM model once we restrict our theory to the
sector of the Fock space over which the \susy\ closes to \Ns
algebra. In section 6,  the relevance of our model to M2-branes is
discussed and  the BPS configurations of our model is analyzed. We
also show that although the theory for a generic configuration is
an \Ns theory, there are BPS configurations for which the theory
can exhibit more fermionic symmetries than is expected from the
\Ns theory. The last section is devoted to summary of our results
and discussions. In the appendix A, we have gathered our
conventions for the $su(N)$ algebras, their representations and
some useful identities among $su(N)$ tensors. In appendix B, we
present the arguments proving that within our setting the extended
three-algebras are only limited to the one generated through
$N\times N$ representation of the $u(N)$ algebra, the
``$u(N)$-based extended three-algebras''. In appendix C, we show
that our $u(2)$-based extended three-algebra is a double cover of
the $so(4)$-based Bagger-Lambert three-algebra. In appendix D, we
show compatibility of the fermionic variations with the $3d$
parity.

\section{Preliminaries of  three-algebras}
In this section we very briefly introduce the notion of
three-algebras and some basic facts about them. We then discuss a
representation of three-brackets of the three-algebras in terms of
four-brackets and ordinary associative algebra of matrices.

\subsection{Introduction to three-algebras}
The three-algebra ${\cal A}_3$ is an algebraic structure defined
through the three-bracket $\bbl\ , \ ,\ \bbr$ \footnote{Since we
will be working with usual matrices and will be using the usual
commutators of matrices and also introduce the new notion of
four-brackets, we will use $\bbl\ , \ ,\ \bbr$ for three-algebra
brackets and usual brackets  for matrix valued objects, either
commutator or
four-brackets.}%
\be\label{3-bracket-def}%
\bbl \Phi_1, \Phi_2, \Phi_3\bbr \in {\cal A}_3, \quad {\rm{for\
any}\ } \Phi_i\in {\cal A}_3, %
\ee%
where%
\be\label{total-anti-smmetry}%
\bbl \Phi_1,\Phi_2,\Phi_3\bbr =- \bbl \Phi_2,\Phi_1,\Phi_3\bbr =-
\bbl \Phi_1,\Phi_3,\Phi_2\bbr=-\bbl \Phi_3,\Phi_2,\Phi_1\bbr%
\ee%
 The three-bracket should satisfy
an analog of the Jacobi identity, the \emph{fundamental identity} \cite{Takhtajan}:%
\be\label{fund-ident}\begin{split}%
{\cal K}_{ij;klm}&\equiv {\bbl} \Phi_i,\Phi_j,\bbl \Phi_k, \Phi_l,
\Phi_m\bbr{\bbr}\cr &={\bbl} \bbl \Phi_i,\Phi_j,\Phi_k\bbr,
\Phi_l, \Phi_m{\bbr}+ {\bbl} \bbl \Phi_i,\Phi_j,\Phi_l\bbr,
\Phi_m, \Phi_k{\bbr}+{\bbl} \bbl \Phi_i,\Phi_j,\Phi_m\bbr,\Phi_k,
\Phi_l{\bbr} .%
\end{split}
\ee%
As we can see ${\cal K}_{ij;klm}$ is anti-symmetric under exchange
of first two  as well as the last three indices.
We equip this algebra with a product $\bullet$ and a Trace%
 \be\label{trace-def}%
 Tr(\Phi_1\bullet \Phi_2)= Tr (\Phi_2\bullet \Phi_1) \in\mathbb{C}
\ee%
with a ``by-part integration'' property%
\be\label{push-thru-trace}%
 Tr(\Phi_1\bullet \bbl\Phi_2, \Phi_3, \Phi_4\bbr )=-
Tr(\bbl\Phi_1,\Phi_2, \Phi_3\bbr \bullet \Phi_4).%
\ee%

$\Phi_i$'s are generically complex valued and we can define the
Hermitian conjugation over the algebra and its three-bracket:%
\be\label{bracket-dagger}%
\bbl \Phi_1, \Phi_2,\Phi_3\bbr^\dagger
=\bbl \Phi_1^\dagger, \Phi_2^\dagger, \Phi_3^\dagger\bbr\ .%
\ee%

Let $T^\alpha$ denote a complete basis in  ${\cal A}_3$, i.e.
$\forall \Phi\in {\cal A}_3,\ \ \Phi=\Phi_\alpha T^\alpha$,
then \eqref{3-bracket-def} implies that%
\be\label{structure-const}%
\bbl T^\alpha, T^\beta,T^\gamma\bbr = f^{\alpha\beta\gamma}_{\quad\
\rho} T^\rho
\ee%
and%
\be\label{metric-def}%
Tr (T^\alpha\bullet T^\beta)\equiv h^{\alpha\beta}
\ee%
defines the metric $h^{\alpha\beta}$ on  ${\cal A}_3$. The metric
$h^{\alpha\beta}$ can in general have positive or negative
eigenvalues, however, $h^{\alpha\beta}$ is always taken to be
non-degenerate and invertible. Noting \eqref{total-anti-smmetry} and
\eqref{push-thru-trace},
\[
f^{\alpha\beta\gamma\delta}\equiv f^{\alpha\beta\gamma}_{\quad\
\lambda} h^{\lambda\delta},
\]
is totally anti-symmetric four-index structure constant. The
fundamental identity in terms of the structure constant $f$ is
written as%
\be\label{fund-iden-f}%
f^{\alpha\beta\gamma}_{\quad\ \lambda}f^{\delta\eta\lambda}_{\quad\
\mu}+f^{\alpha\beta\delta}_{\quad\
\lambda}f^{\eta\gamma\lambda}_{\quad\
\mu}+f^{\alpha\beta\eta}_{\quad\
\lambda}f^{\gamma\delta\lambda}_{\quad\ \mu}
=f^{\gamma\delta\eta}_{\quad\ \lambda}f^{\alpha\beta\lambda}_{\quad\ \mu}.%
\ee%

It has been shown that \cite{Papadopoulos:2008sk} for Euclidean
case, when $h^{\alpha\beta}$ is positive definite,
\eqref{fund-iden-f} has only a single solution
$f^{\alpha\beta\gamma\delta}\propto
\epsilon^{\alpha\beta\gamma\delta}$, while when $h^{\alpha\beta}$ is
Lorentzian (when $h$ has a single negative eigenvalue), one can
associate a three-algebra structure to any  Lie-algebra \cite{Russo,
Verlinde, HIM}.  In this case the fundamental identity reduces to
the Jacobi identity of the algebra and the structure constant of the
\ta\ is expressed in terms of the structure constant of the
underlying Lie-algebra.

We would like to comment that for the Euclidean and the Lorentzian
cases one can choose a Hermitian basis $T^\alpha$ for which the
structure constants $f^{\alpha\beta\gamma\delta}$ are real valued.

\subsection{Four-bracket representation for three-algebras}\label{4b-rep-3algebras-section}

As discussed in \cite{ASS} one may give a representation of \ta s
in terms of ordinary algebra of matrices. To that end  we need to
give a four-bracket realization for the three-brackets of the
three-algebra:%
\be\label{threeVsfour}%
\bbl A_1, A_2, A_3\bbr \equiv [ \hat A_1, \hat A_2, \hat A_3, T] %
\ee%
where the hatted quantities are just normal matrices and $T$ is a
matrix which anticommutes with all the other elements of the algebra%
\be\label{T-anticommute}%
\{A_i, T\}=0\ . %
\ee%
The four-bracket is defined as antisymmetrized product of the
elements appearing inside, that is%
\be\label{four-bracket-def}\begin{split}%
\hspace*{-0.5cm} [\hat A_1, \hat A_2, \hat A_3, \hat A_4] &
=\frac{1}{4!} \epsilon^{ijkl} \hat A_i \hat A_j \hat A_k \hat A_l
\cr &
 =\frac{1}{4!}\left(\{[\hat A_1, \hat A_2], [\hat A_3, \hat A_4]\}-\{[\hat A_1, \hat A_3], [\hat A_2, \hat
 A_4]\}+\{[\hat A_1, \hat A_4], [\hat A_2, \hat A_3]\}\right).
\end{split} %
\ee%
The fundamental identity \eqref{fund-ident} in terms of the
four-bracket takes the form\footnote{Hereafter we will drop the hats
on any matrix $A$.}
\be\label{fund-iden-4bracket}\begin{split}%
[[A_1, A_2,B_1, T], B_2, B_3 ,T]&+[ B_1,[A_1,A_2, B_2, T],B_3, T]
\cr
& +[B_1,B_2,[A_1,A_2,B_3,T],T]=[A_1,A_2, [B_1,B_2,B_3, T],T], %
\end{split}\ee%
for any element $A_i$ and $B_i$ in the algebra. Working with
matrices, we can choose the trace over the matrices as the natural
trace over our \ta .

 It is evident that with the above definitions
not all arbitrary sets of matrices satisfy the closure
\eqref{3-bracket-def} and fundamental identity
\eqref{fund-iden-4bracket}. It is, however, immediate to check that
 within our matrix representation and the four-bracket, the trace
condition \eqref{push-thru-trace} and the Hermitian conjugation
\eqref{bracket-dagger} (if $T=T^\dagger$) are automatically
satisfied. In \cite{ASS} it was shown that the only set of
matrices which satisfy the closure and fundamental identity
requirements as stated above, are the ``$so(4)$-based'' algebras
(where $A_i$'s and $T$ are respectively taken to be $N\times N$
representation of $so(4)$ Dirac $\gamma$-matrices and the
$\gamma^5$), compatible with the three-algebra no-go theorem
\cite{Papadopoulos:2008sk}.

\section{$u(N)$-based {extended} three-algebras}

As was argued by Bagger and Lambert \cite{BL2} the requirement of
fundamental identity for the three-algebras is demanded by the
``gauge symmetry'' as well as the closure of the supersymmetry
algebra in the BLG theory. The ``Tr'' operation (and hence the
metric), however, is needed to construct ``gauge invariant''
physical observables. Given the restrictions on the construction of
the \ta s one is hence motivated to see if the notion of \FI\ and/or
the closure condition can be relaxed or extended in such a way that
the gauge invariance and the \Ne supersymmetry algebra requirements
are met, while allowing for further possibilities of \ta s.

In \cite{ASS} one such possibility, which were dubbed as the
\emph{relaxed \ta s}, was explored. There, it was noted that by
the addition of a ``spurious'' part of the algebra of matrices one
can relax the closure condition and the \FI\ holds up to the
``spurious'' parts, while keeping the virtues resulting from those
properties. In this way an explicit matrix representation for the
Lorentzian \ta s were given and was shown that the Lorentzian \ta\
is a unique outcome of the non-empty spurious part of the algebra
\cite{ASS}.

Here we study yet another way of extending the notion of the \ta s
by revisiting the notion of the \FI . As it will become clear in
the next sections, what is  needed to ensure the gauge symmetry
closure is not the strict form of the \FI\ given in
\eqref{fund-ident} or \eqref{fund-iden-4bracket}. A similar
observation has also been made in \cite{BL3-bracket-N6}. In
\cite{BL3-bracket-N6}, however, the focus was working with
non-totally antisymmetric three-brackets, whereas in our case the
brackets are still totally antisymmetric and the implementation of
the \FI\ is modified. This will become clear in this section.

In what follows based on the appropriate notion of extended \FI ,
we construct the \emph{extended \ta }, using our four-bracket and
matrix representation introduced in the previous subsection.

\subsection{Construction of the extended three-algebras}%

To start we assume that the complete basis for the \ta \ is of the
following form%
\be\label{TM-basis}
 T^M\in \{T^A_+,\ T^A_-,\ T\}\ ,
\ee%
with%
\be\label{TApm-T}%
T^A_{\pm}= t^A \otimes \sigma^\pm, \qquad  T=\id_N\otimes \sigma^3\ , %
\ee%
where $t^A$ are (yet to be specified) set of $N\times N$
\emph{Hermitian} matrices and $\sigma^{\pm},\ \sigma^3$ are the
$2\times 2$ Pauli matrices%
\be\label{Pauli-sigma}%
[\sigma^+, \sigma^-]=\sigma^3,\quad  [\sigma^3, \sigma^\pm]=\pm
2\sigma^\pm,\quad  \{\sigma^+, \sigma^-\}=\id_{2\times 2} . %
\ee%
Since $t^A$'s are Hermitian,%
\be\label{TA-dagger}%
(T^A_+)^\dagger=T^A_-\ .%
\ee%
With the above it is clear that%
\be\label{TA-T}%
\{T^A_\pm, T\}= 0\ , \qquad T^2=\id_{2N\times 2N},\quad [T,
T^A_{\pm}]=\pm 2T^A_\pm\ , %
\ee%
moreover,%
\be\label{Tpm2=0}%
T^A_+ T^B_+= T^A_- T^B_-=0\ . %
\ee%
We normalize our basis such that%
\be\label{TA-normalization}%
Tr(T^A_+ T^B_-)= Tr(T^A_- T^B_+)=\half \delta^{AB}\ .%
\ee%

Let us consider the most general four-bracket $[T^M, T^N, T^P,
T]$. It is evident that if any of $T^M, T^N$ or $T^P$ is $T$ the
bracket vanishes. We hence remain with four types of
four-brackets, two of them are those which only involve   $T^A_+$
or
$T^A_-$ identically vanish, %
\be\label{three-plus-or-minus}%
[T^A_+, T^B_+, T^C_+, T]=[T^A_-, T^B_-, T^C_-, T]=0 , %
\ee%
where we have used $(\sigma^+)^2=(\sigma^-)^2=0$ and the definition
of the four-bracket. The other two are those with two $T^A_+$ and
one $T^A_-$ or two $T^A_-$ and one $T^A_+$, which are related by
Hermitian conjugation%
\be\label{4bracket-conj}%
\left([T^A_+, T^B_-, T^C_+, T]\right)^\dagger =[T^A_-, T^B_+, T^C_-, T]\ ,%
\ee%
where we have used \eqref{TA-dagger}. Therefore there is only a
single type of independent four-bracket.

Using straightforward algebra of Pauli matrices and the definition
of the four-bracket we have%
\be\label{bracket-computed}%
[T^A_+, T^B_-, T^C_+, T]=\frac{-1}{6}\left(t^A t^B t^C - t^C t^B
t^A\right)\otimes \sigma^+\ . %
\ee%

\subsection{Closure condition}

Demanding the closure of the four-bracket over the set of $T^A_+$
and $T^A_-$ requires that%
\be\label{closure-tA}%
\frac{-1}{6}\left(t^A t^B t^C - t^C t^B t^A\right)=f^{ABC}{}_{D}\
t^D%
\ee%
for some numeric coefficients $f^{ABC}{}_D$. If we choose to work
with $t^A$ which are generators of a (semi-simple) Lie-algebra,
\footnote{It is worth noting that this is a working assumption and
 not a necessary one.} the above closure condition
\eqref{closure-tA} is very restrictive and uniquely fixes this
algebra to be a $u(N)$ (for arbitrary $N$). Moreover, it also
requires $t^A$'s to be in the $N\times N$ fundamental
representation of the $u(N)$ algebra. In other words, the closure
condition \eqref{closure-tA} is only satisfied for the algebras
which are their own enveloping algebra and $u(N)$ in the $N\times
N$ representation is the only such algebra. In the appendix B, we
present a proof of this statement. These algebras will hence be
called $u(N)$-based (extended) three-algebras. Using
\eqref{closure-tA} we have%
\be\label{++-T}%
\begin{split}%
[T^A_+, T^B_-, T^C_+, T] &=f^{ABC}{}_D T^D_+\ ,\cr %
[T^A_-, T^B_+, T^C_-, T] &=-f^{ABC}{}_D T^D_-\ .
\end{split}%
\ee%
In the second identity we have used the  fact that, noting
\eqref{closure-tA} and hermiticity of $t^A$'s, $f$ is pure
imaginary.

Using \eqref{push-thru-trace} we have%
\be\label{f-all-up}%
f^{ABCD}=-2\ Tr\left([T^A_+, T^B_-, T^C_+, T^D_-]T\right)\ . %
\ee%
The above explicitly shows that%
\be\label{f-symmetries}%
f^{ABCD}=-f^{CBAD}=-f^{ADCB}=+f^{CDAB}=-f^{BADC}=-(f^{ABCD})^*\ .
\ee%
For the last two identities we have used the fact that $f$ is pure
imaginary. From \eqref{bracket-computed} and that $t^A t^A\propto \id$, it is readily seen that%
\[
\sum_A\ f^{AABC}=0\ .
\]

We would like to comment that  $f^{ABCD}$ with the above symmetry
properties may be viewed as the structure constant of a new type
(or ``generalized'') three-algebra \cite{BL3-bracket-N6,Cherkis,
JoseF-O}. The three-bracket of these generalized three-algebras
are hence not totally antisymmetric and as a consequence their
fundamental identity is expressed in a bit different way than
\eqref{fund-ident}. Our notion and realization of the extended
three-algebras, although looking similar to the constructions
discussed \cite{BL3-bracket-N6,Cherkis, JoseF-O}, has its own
specific features. In particular, as is explicitly seen from the
definition of our brackets \eqref{threeVsfour} and
\eqref{four-bracket-def}, our four-brackets are  antisymmetric
under exchange of any two elements. Therefore, in the $M, N, P$
basis and before expansion in $T^A_{\pm},\ T$ basis, the structure
constant $\hat f$,
\[{\hat f}^{MNPQ}\equiv -Tr([T^M, T^N, T^P, T^Q] T), \]
is totally antisymmetric. Moreover, we have an explicit matrix
representation and $u(N)$ algebra has a distinguished role in our
setting.

For the specific choice of $u(N)$ basis   given in the appendix A
(where $t^a$'s are generators of $su(N)$ part of $u(N)$ and
$t^0\propto \id$ is its $u(1)$ part) one can show that:%
\bse\label{f-explicit}%
\begin{align}%
f^{00ab}&=0, \\%
f^{0abc}&= f^{a0bc}=f^{ab0c}=f^{abc0}=\frac{-i}{6}\cdot
\frac{1}{\sqrt{2N}}\ f^{abc},\\ %
f^{abcd}&=\frac{-i}{12}\left(f^{abe}d^{cde}+f^{cde}d^{abe}\right).%
\end{align}%
\ese%

It is worth noting that for the specific case of $N=2$, the $u(2)$
algebra, $d^{abc}=0$ and hence $f^{abcd}=0$. In this case the only
non-vanishing components of $f$ are $f^{0abc}\propto
\epsilon^{abc},\ a,b,c=1,2,3$. As it has been shown in appendix B,
for the $N=2$ case one can choose a sector (by working with half of
the eight $T^A_\pm$ generators) in which the structure constants
become totally antisymmetric. Among the $u(N)$ based (extended)
three-algebras the $u(2)$ case is the only one with the possibility
of totally antisymmetric structure constant.

\subsection{Extended \FI}

As discussed (e.g. see \cite{BL2}) the \FI\ \eqref{fund-ident} or in
its four-bracket presentation \eqref{fund-iden-4bracket} is
necessitated by the gauge invariance and the superalgebra closure of
the BLG theory. However, as will become clear in the next section,
these conditions might be met through a bit weaker condition than
\eqref{fund-iden-4bracket}: It is enough to check the \FI\
\eqref{fund-iden-4bracket} for the case when either of $A_1, A_2$
are of the form of $T^A_+$ and $T^A_-$ (and not both of them of the
form of $T^A_+$ or $T^A_-$) while $B_i$'s can be arbitrary. In terms
of our basis that is,
\be\label{FI-extended}\begin{split}%
[[T^A_+, T^B_-,T^M, T], T^N, T^P ,T]&+[ T^M,[T^A_+, T^B_-, T^N,
T],T^P, T] \cr
& +[T^M,T^N,[T^A_+, T^B_-,T^P,T],T]=[T^A_+, T^B_-, [T^M,T^N,T^P, T],T], %
\end{split}\ee%
where $T^M, T^N, T^P$ are either $T^A_+$, $T^A_-$ or $T$.

Recalling  the discussions of sections 3.1 and 3.2, the extended
\FI\ \eqref{FI-extended} for  $(T^M, T^N, T^P)=(T^C_+, T^D_+,
T^E_+)$ or $(T^C_-, T^D_-, T^E_-)$ is trivially satisfied while it
should be checked for $(T^M, T^N, T^P)=(T^C_+, T^D_+, T^E_-)$ or
$(T^M, T^N, T^P)=(T^C_+, T^D_-, T^E_-)$ (or in general two plus
and a minus or two minus and a plus type generators) cases. These
two cases, however, are not independent and are related by complex
conjugation. Therefore, we will only need to verify one of these
cases which we choose it to be $(T^M, T^N, T^P)=(T^C_+, T^D_-,
T^E_+)$. It is straightforward to verify that \FI\
\eqref{FI-extended} is satisfied for this case. This may be done
directly using \eqref{bracket-computed} and the associativity of
the product of $t^A$'s (without using the fact that $t^A$'s are
generators of $u(N)$). Since, as discussed in section 3.2, the
closure condition requires that in our extended three-algebras
$t^A$'s must be generators of $u(N)$, we call them
\emph{$u(N)$-based extended three-algebras}.

It is useful to represent the fundamental
identity in terms of the ``structure constants'' $f^{ABCD}$:%
\be\label{FI-fABCD}%
f^{ABGH}\ f^{CDFG}+ f^{ABGD}\ f^{CGFH}+f^{ABCG}\
f^{FDGH}=f^{ABFG}\ f^{CDGH}\ . %
\ee%
Note that the indices on $f$ are lowered and raised by the metric
defined in \eqref{TA-normalization}, i.e. $\delta_{AB}$ when we work
with  $A$ and $B$ indices instead of $M$ and $N$ indices. One can
also  verify that the above identity is fulfilled using the explicit
expression for $f$ given in \eqref{f-explicit} and using the
identities given in the appendix A. In the appendix B we show the
connection between the Bagger-Lambert three-algebra and the
$u(2)$-based extended three-algebra.

\section{The $SO(8)$ invariant SCFT action}

Since the  on-shell matter content of the \3dNe SCFT should involve
eight \emph{real valued} three-dimensional scalars $X^I$,
$I=1,2,\cdots ,8$ in the $\mathbf{8_v}$ of the $SO(8)$ R-symmetry
group, eight two component Majorana (real valued) three-dimensional
fermions $\Psi$ (i.e. they satisfy $\gamma^{012}\Psi=\Psi$) in the
$\mathbf{8_s}$ of SO(8), we start with this explicitly $SO(8)$
notation. Unless there can be confusion, here we will suppress both
the $3d$ and the R-symmetry fermionic indices. Each of the above
physical fields, which will generically be denoted by $\Phi$, are
also
assumed to be elements of the $u(N)$-based extended three-algebra and hence%
\be\label{phys-fields}%
\Phi= \Phi_M T^M= \Phi_A^+\ T^A_++ \Phi_A^-\ T^A_-+\Phi_T\ T.%
\ee%

 As argued by Bagger and Lambert \cite{BL2} and Gustavson \cite{Gust1} to close the
 $\cN>4$
supersymmetry algebra, besides the above propagating physical
fields we need to introduce a non-propagating gauge field with a
\cs action. The gauge field should have two three-algebra indices,
i.e.
\be\label{gauge-field}%
A_{\mu}= \half A_{\mu AB}\ [T^A_+, T^B_-]\ .
\ee%
We would like to emphasize that the $A_{\mu AB}$ components are not
anti-symmetric under the exchange of $A$ and $B$ indices.

As we will show in this section, the three-algebra with the
extended notion of the \FI\ \eqref{FI-extended} is enough to
ensure the closure of the gauge transformations. The extended \FI,
however, is not enough to guarantee the closure of the
\emph{$SO(8)$ covariant} (i.e. \Ne ) supersymmetry
transformations. As a result we are forced to close the
supersymmetry onto a smaller set of states. As we will show the
largest set of such states keep $SU(4)\simeq SO(6) \in SO(8)$
(i.e. ${\cal N}=6$) supersymmetry.

\subsection{ The BLG Lagrangian in terms of four-brackets}

As discussed in \cite{ASS} one can represent  the BLG theory in
terms of the four-brackets. This representation explicitly
exhibits the $SO(8)$ invariance of the theory. Here we take the
physical fields and the four-brackets to be in the $u(N)$-based
extended three-algebra discussed in the previous section.
\paragraph{The gauge invariant action with explicit $SO(8)$ symmetry}
\be\label{BLG-action-four-bracket} \begin{split}%
  S&=\int d^3x\ Tr\biggl[-\half D_\mu
  X^ID^\mu X^I-\frac{1}{2.3!}[X^I,X^J,X^K,T][X^I,X^J,X^K,T]\cr
 &+\frac{i}{2}\bar{\Psi}\gamma^\mu
  D_\mu\Psi-\frac{i}{4}[\bar{\Psi},X^I,X^J,T] \Gamma^{IJ}\Psi\cr%
   &+\frac{1}{2}\epsilon^{\mu\nu\rho}\left(A_{\mu AB}\partial_\nu A_{\rho
CD} T^D_-+\frac{2}{3} A_{\mu AB} A_{\nu CD} A_{\rho EF} [T^D_-,
T^E_+, T^F_-, T]\right) [T^A_+, T^B_-, T^C_+, T]
\biggr],%
\end{split}\ee %
where the trace is over $2N\times 2N$ matrices and%
\be\label{cov-der}%
D_\mu\Phi \equiv \partial_\mu \Phi - {A}_{\mu AB}  [T^A_+,
T^B_-, \Phi, T]\ .%
\ee%
In terms of the components it is
\bse\label{cov-der-comp}%
\begin{align}%
(D_\mu\Phi)_T &= \partial_\mu \Phi_T  \\%
(D_\mu\Phi)_D^+ &= \partial_\mu \Phi_D^+ - f^{ABC}{}_{D} {A}_{\mu
AB}\ \Phi_C^+ \\%
(D_\mu\Phi)_D^- &= \partial_\mu \Phi_D^- + f^{ABC}{}_{D} {A}_{\mu
BA}\ \Phi_C^-\ ,%
\end{align}%
\ese%
where in (\ref{cov-der-comp}c) we have used the properties of
$f^{ABCD}$ \eqref{f-symmetries}.

With the above definition it is seen that if $\Phi=\Phi^\dagger$,
then $D_\mu\Phi=(D_\mu\Phi)^\dagger$. Moreover,%
\be\label{A-hermiticity}%
A_{\mu AB}^*=-A_{\mu BA}\ , %
\ee%
where $*$ is the complex conjugation. In terms of the gauge field
$A_\mu$ \eqref{gauge-field}, i.e. $A_\mu^\dagger=-A_\mu$.
As in \cite{BL2} it is useful to define a new gauge field%
\be\label{tildeA}%
{\tilde A}_{\mu CD}=f^{ABCD} A_{\mu AB}\ . %
\ee%
In terms of $\tilde A_\mu$ the covariant derivatives take the form
\be\label{cov-der-tildeA}%
(D_\mu\Phi)_A^+ = \partial_\mu \Phi_A^+ -  \tilde{A}_{\mu BA}\
\Phi_B^+,\qquad %
(D_\mu\Phi)_A^- = \partial_\mu \Phi_A^- + \tilde{A}_{\mu AB}\ \Phi_B^- .%
\ee%
It is worth noting that the $\tilde A_{\mu}$ gauge field,
similarly to $A_{\mu AB}$, has only $[T^A_+, T^B_-]$ components.

\paragraph{Gauge transformations}%
\bse\label{gauge-sym}\begin{align} %
\delta_{gauge} \Phi_T &=0 \\ %
\delta_{gauge} \Phi_A^+&= \tilde{\Lambda}_{BA}\Phi_B^+\ , \qquad
\delta_{gauge} \Phi_A^-= -\tilde{\Lambda}_{AB}\Phi_B^-\ , \\%
\delta_{gauge} {\tilde A}_{\mu AB} &= \partial_\mu
{\tilde\Lambda}_{AB}+ \left( {\tilde A}_{\mu AC}{\tilde
\Lambda}_{CB}-{\tilde \Lambda}_{AC}{\tilde A}_{\mu CB}\right)\ .%
\end{align}
\ese%
Note that like the $\tilde A_\mu$, $\tilde\Lambda$ has only
components along $[T^A_+, T^B_-]$.

From the
above it is readily seen that%
\be%
\delta_{gauge}(D_\mu\Phi)_A^+=\tilde{\Lambda}_{BA}(D_\mu\Phi)_B^+\ , %
\qquad%
\delta_{gauge}(D_\mu\Phi)_A^-=-\tilde{\Lambda}_{AB}(D_\mu\Phi)_B^-\ .%
\ee%

The action
\eqref{BLG-action-four-bracket} is invariant under  the above
gauge transformations provided that%
\be\label{gauge-closure}%
\delta_{gauge}\left([\Phi_1, \Phi_2, \Phi_3,
T]\right)=[\delta_{gauge}\Phi_1, \Phi_2, \Phi_3, T]+[\Phi_1,
\delta_{gauge}\Phi_2, \Phi_3, T]+[\Phi_1, \Phi_2,\delta_{gauge}
\Phi_3, T].%
\ee%
This identity  holds as a result of the extended \FI \
\eqref{FI-extended}, once we recall that the gauge transformations
parameter $\Lambda$ has one plus type and one minus type $T^A$
generators. As a result of the extended \FI\ one can also show that%
\be\label{cov-bracket}%
D_\mu\left([\Phi_1, \Phi_2, \Phi_3, T]\right)=[D_\mu \Phi_1, \Phi_2,
\Phi_3, T]+[\Phi_1, D_\mu\Phi_2, \Phi_3, T]+[\Phi_1, \Phi_2,D_\mu
\Phi_3, T].%
\ee%
Eqs.\eqref{gauge-closure} and \eqref{cov-bracket} are nothing but
the statement of closure of the gauge symmetry algebra  of the
action \eqref{BLG-action-four-bracket}.

So far we have presented a theory which enjoys  the gauge symmetry
\eqref{gauge-sym} as well as global $SO(8)$ and $3d$ Poincar\'e
invariance. The propagating bosonic degrees of freedom of this
theory are $X^I_T$, $(X^I)^+_A$, $(X^I)^-_A$. $X^I_T$ are eight real
\emph{free} scalars which decouple from the rest of the theory. The
$X^I_T$ piece, together with its fermionic counterpart $\Psi_T$ form
a trivial \Ne superconformal theory (with the explicit supersymmetry
transformation given in the next subsection). Hereafter, we will
hence ignore the $\Phi_T$ piece by simply setting them to zero.
$(X^I)^+_A=\left((X^I)^-_A)\right)^*$ which are elements of $N\times
N$ matrices for the $u(N)$-based algebra, parameterize $8N^2$
complex (or $8\cdot 2N^2$ real) scalars. However, the \Ne theory is
expected to have real valued scalars. As we will see the closure of
the supersymmetry and parity invariance of the physical Fock space
of the theory should be used to reduce this extra degrees of
freedom.

\subsection{Parity invariance}

The \3dNe theory is expected to be  invariant under the $3d$ parity
transformations $x^0,x^1\rightarrow x^0, x^1$ and $x^2\rightarrow
-x^2$. The parity invariance of the (twisted) \cs term implies that
under parity
\be\label{parity-gauge-field} %
{\tilde A}_{0 AB}, {\tilde A}_{1 AB}\longrightarrow -{\tilde A}_{0
BA}, -{\tilde A}_{1 BA}, \qquad {\tilde A}_{2 AB}\longrightarrow
+{\tilde A}_{2 BA}\ .
\ee%
Recalling \eqref{A-hermiticity}, that is%
\be\label{parity-gauge-field-cc}%
{\tilde A}_{\mu AB}\stackrel{\rm parity}
{\longleftarrow\!\!\longrightarrow} ({\tilde A}^p_{\mu AB})^*\ ,
\ee%
where by $A^p_{\mu AB}$ we mean a vector with components $A_{0
AB},\ A_{1 AB},\ -A_{2 AB}$.

 The parity invariance of the kinetic terms, as well as
the interaction terms imply that under parity one should exchange
the
plus and minus components, for the scalar fields that is,%
\be\label{parity-scalar}%
(X^I)_A^+ \stackrel{\rm parity}
{\longleftarrow\!\!\longrightarrow} (X^I)_A^-\ ,%
\ee%
and for $3d$ fermions%
\be\label{parity-fermions}%
\Psi_A^+ \stackrel{\rm parity} {\longleftarrow\!\!\longrightarrow}
\gamma^2 \Psi_A^- \ . %
\ee%
\eqref{parity-gauge-field}, \eqref{parity-scalar} and
\eqref{parity-fermions} can be combined into the fact that under
parity $T^A_+\leftarrow\!\rightarrow T^A_-$,  $T \rightarrow -T$. It
is useful to introduce action of the parity on the
$X^I$, $\Psi$ and $A_\mu$ fields:%
\be\label{XI-Psi-Parity}%
\begin{split}%
(X^I)_{parity} &=(X^I)^-_A\ T^A_+ + (X^I)^+_A\ T^A_- 
\cr
(\Psi)_{parity}&=\gamma^2\Psi^-_A\  T^A_+ +
\gamma^2\Psi^+_A\  T^A_- 
\cr%
(A_{\mu})_{parity}&= \frac12 A^p_{\mu AB} [T^A_-, T^B_+]\ .%
\end{split}%
\ee%
(Note that, as discussed earlier, we have set the $X_T$ and $\Psi_T$
components to zero.)
Using the above and \eqref{f-symmetries} one can show that%
\be\label{bracket-parity}%
\left([\Phi_1, \Phi_2,\Phi_3,
T]\right)_{parity}=-[(\Phi_1)_{parity},
(\Phi_2)_{parity},(\Phi_3)_{parity}, T]\ , %
\ee%
where $\Phi_i$ are either $X^I$ or $\Psi$. With these and noting
that $\bar\Psi\Psi$ is a pseudoscalar \cite{schwarz} one can show
that the action \eqref{BLG-action-four-bracket} is invariant under
parity. Although the action \eqref{BLG-action-four-bracket} is
parity invariant, the physical fields $X^I$ in general are not.

We  point out that if under parity the gauge parameter $\tilde
\Lambda_{AB}$ transforms as $\tilde \Lambda_{AB}\to
-{\tilde\Lambda}_{BA}$, the gauge transformations \eqref{gauge-sym}
are compatible with the parity.
  As discussed, among the gauge field components ${\tilde
A}_{\mu
AB}$, the antisymmetric part %
\be%
{\tilde A} _{\mu [AB]}=\half(\tilde A_{\mu AB}-\tilde A_{\mu BA})%
\ee%
transforms as a vector, and the symmetric part%
\be%
\tilde A_{\mu \{AB\}}=\half(\tilde A_{\mu AB}+\tilde A_{\mu BA})\ ,%
\ee%
transforms as a pseudovector.

It is worth noting that, as can be seen from \eqref{phys-fields} and
\eqref{gauge-field}, the action \eqref{BLG-action-four-bracket} is
invariant under another global $U(1)$ symmetry, the $U(1)_\lambda$
symmetry: $T^A_\pm \longrightarrow e^{\mp i\lambda} T^A_\mp$, while
keeping $\Phi$ \eqref{phys-fields} and  $A_\mu$ \eqref{gauge-field}
invariant, explicitly that is,%
\be\label{global-U(1)b}%
\Phi_A^\pm\longrightarrow e^{\pm i\lambda} \Phi_A^\pm\ ,\qquad
\Phi_T\rightarrow \Phi_T\ , \qquad A_{\mu AB}\rightarrow A_{\mu AB} \ .%
\ee%
The parity changes the sign of the charge under the $U(1)_\lambda$
symmetry. We will comment on $U(1)_\lambda$ further in sections 5
and 6. We also note that $\sigma^\pm,\ \sigma^3$ form an $su(2)$
algebra and the $U(1)_\lambda$ and parity are forming an $O(2)$
automorphism of this $su(2)$ algebra.

\subsection{Supersymmetry transformations and their closure }

After discussing the gauge and parity invariance of  our theory,
we now discuss its  supersymmetry. Since the action
\eqref{BLG-action-four-bracket} is essentially the Bagger-Lambert
action \cite{BL2}, and recalling that our four-brackets are
totally antisymmetric with the trace property
\eqref{push-thru-trace}, we propose the
following fermionic (or supersymmetry) transformations%
\bse\label{SUSY-trans-four-bracket}\begin{align} %
 \delta X^I&=i\bar{\epsilon}\Gamma^I\Psi \\ %
 \delta\Psi&=D_\mu X^I\Gamma^I\gamma^\mu\epsilon-\frac{1}{6}[X^I,X^J,X^K,T]
 \Gamma^{IJK} \epsilon \\%
\delta \tilde A_{\mu AB} &=if_{ABCD}\
\bar{\epsilon}\gamma_\mu\Gamma^I\left((X^I)_C^+\Psi^-_D-(X^I)_D^-\Psi^+_C\right) .%
\end{align}%
\ese %
The fermionic transformation parameter $\epsilon$ is a $3d$
anti-Majorana fermion
\be\label{epsilon-fermion}%
\gamma^{012}\epsilon=-\epsilon\ ,%
\ee%
and is in $\mathbf{8}_c$ of $SO(8)$ (in contrast with $\Psi$ which
is in $\mathbf{8}_s$).

As first step we check if the above transformations keep the action
\eqref{BLG-action-four-bracket} invariant. The
variation of the action under the above transformations is%
\be\label{susy-action-var}%
\begin{split}%
\delta S &= \int d^3x\ Tr\left(E.o.M_{X^I}\ \delta
X^I+E.o.M_{\Psi}\ \delta\Psi\right)+E.o.M_{A_{\mu AB}}\ \delta
A_{\mu AB}+
\partial_\mu J^\mu\ ,\cr
J^\mu &= Tr \left(-D^\mu X^I \delta X^I+ i{\bar \Psi}\gamma^\mu
\delta \Psi+ 
\epsilon^{\mu\nu\alpha} {A}_{\nu}\delta
{\tilde A}_{\alpha}\right)\ ,
\end{split}%
\ee%
where the first three terms vanish on the solutions of equations of
motion and $J^\mu$ after some algebraic manipulations takes the form%
\be\label{susy-var-Jmu}%
J^\mu=i\bar\epsilon
\left(
-\gamma^{\mu\nu}{\tilde A}_{\nu
CD}\Gamma^K\left((X^K)_C^+\Psi^-_D-(X^K)_D^-\Psi^+_C\right)-\frac{1}{6}\gamma^\mu
Tr([X^I,X^J,X^K,T]\Psi)\Gamma^{IJK}\right)
.
\ee%
For the invariance of the action $\partial_\mu J^\mu$ must vanish
for any arbitrary $\epsilon$. This can, however, happen in a
specific gauge. It is straightforward to check that if $f^{ABCD}$
were totally antisymmetric then in the gauge $2\gamma^{\nu}{\tilde
A}_{\nu
AD}=
3f_{ABCD} \Gamma^{IJ} (X^I)^-_B (X^J)^+_C, $ %
$\partial_\mu J^\mu$ would vanish when sandwiched between any two
$\epsilon$-type (i.e $3d$ anti-Majorana and in $\mathbf{8}_c$ of
$SO(8)$) fermions.  For our case, however, $f^{ABCD}$ is not totally
anti-symmetric and in the above gauge $\partial_\mu J^\mu$ does not
vanish.\footnote{The point that with $f^{ABCD}$ which is not totally
antisymmetric we cannot keep $16$ supersymmetries were mentioned in
\cite{BL3-bracket-N6} and further emphasized to us by N. Lambert.}
As will become clear momentarily we choose
to work in the gauge where%
\be\label{SUSY-gauge}%
\gamma^{\nu}{\tilde A}_{\nu
AB}=
+f_{ACDB} \Gamma^{IJ} (X^I)^-_C (X^J)^+_D, %
\ee%
when sandwiched between any two $\epsilon$-type fermions.
 In this gauge we have
\be%
\delta S=\int \partial_\mu\biggl(i\bar\epsilon\gamma^\mu
\bigl(\Gamma^I (X^I)^-_A\Psi^-_B \chi^+_{AB}- \Gamma^I
(X^I)^+_A\Psi^+_B
\chi^-_{AB}\bigr)\biggr)%
 \ee%
where%
 \be\label{chipm-def}%
{\chi}^+_{AB} \equiv f_{ACBD}\ \Gamma^{JK}\ (X^J)^+_C\ (X^K)^+_D\
,\qquad {\chi}^-_{AB} \equiv f_{ACBD}\ \Gamma^{JK}\ (X^J)^-_C\
(X^K)^-_D\ . %
\ee%
Invariance of the action then demands that $\chi^\pm=0$. As we will
see closure of the fermionic transformations onto the $3d$
super-Poincar\'e algebra again demands vanishing of $\chi^\pm$, the
condition which will be satisfied for a specific subset of fermionic
transformations once the degrees of freedom are also restricted to
certain subsector of $SO(8)$ states.

\subsubsection{Closure of supersymmetry algebra}

As a parallel but equivalent analysis, we also study the closure of
two successive fermionic transformations on the fields in our
action. The closure of the (on-shell) \Ne (that is, 16 on-shell
supersymmetries) demands that two successive \susy\ transformations
of $X^I$, $\Psi$ and the gauge field $A_{\mu AB}$, up to gauge
transformation  and upon using the equations of motion, on the
\emph{physical Fock space of the theory} must close onto the $3d$
Poincar\'e \cite{BL2}.

Our \susy\ transformations are formally the same as those introduced
in \cite{BL2} and \cite{BL3-bracket-N6}, once they are represented
in terms of three-brackets, two successive \susy\ transformations
lead to the same results as in \cite{BL2, BL3-bracket-N6} and most
of the analysis are the same as those appeared in \cite{BL2,
BL3-bracket-N6}. Therefore we do not present the details of the
computations and only stress the points of difference. Three closure
conditions should be verified: \footnote{We would like to thank Neil
Lambert for his fruitful and critical comments on the closure of
\susy\ in our model.}

$\bullet$ Closing the \susy\ on the scalars we find \cite{BL2}%
\be\label{susy-closing-XI}%
[\delta_1,\delta_2] X^I=v^\mu D_\mu X^I-V_{JK}[X^I,
X^J, X^K, T]\ , %
\ee%
where \be\label{vmu-VJK}%
v^\mu=-2i\bar \epsilon_2\gamma^\mu
\epsilon_1,\qquad
V_{JK}=-i\bar\epsilon_2\Gamma_{JK}\epsilon_1\ . %
\ee%
Let us now consider the $T^A_+$ and $T^A_-$ components. We note that
the $T^A_+$ component of $V_{JK}[X^I, X^J, X^K, T]$ involves both
$(X^I)_B^+$ and $(X^I)_B^-$ components, while the $T^A_+$ component
of $D_\mu X^I$ is only involving $(X^I)^A_+$ (\emph{cf.}
\eqref{cov-der-comp}). \footnote{Although very similar our case, the
extra term proportional to $(X^I)^\pm$ in the variation of
$(X^I)^\mp$ do not happen in the analysis of \cite{BL3-bracket-N6}
because, unlike ours,
their bracket is not totally anti-symmetric.} Explicitly,%
\be\label{susy-closing-XI-component-1}%
\begin{split}%
[\delta_1,\delta_2] (X^I)^+_D &=v^\mu \partial_\mu (X^I)^+_D+
\bigl(\tilde{\Lambda}_{AD} -v^\mu {\tilde A}_{\mu AD}\bigr)
(X^I)^+_A-i\bar\epsilon_2\chi_{AD}^+\epsilon_1\ (X^I)^-_A\cr%
[\delta_1,\delta_2] (X^I)^-_D &=v^\mu \partial_\mu (X^I)^-_D-
\bigl(\tilde{\Lambda}_{DA} -v^\mu {\tilde A}_{\mu DA}\bigr)
(X^I)^-_A+i\bar\epsilon_2\chi_{AD}^-\epsilon_1\ (X^I)^+_A,
\end{split}%
\ee%
where%
\be\label{Lambda}%
\tilde{\Lambda}_{AD} \equiv 2f_{ABCD}\ V_{JK}\ (X^J)^-_B
(X^K)^+_C\ ,
\ee%
and $\chi^\pm$ are defined in \eqref{chipm-def}.

 Due to the presence of the $\chi$ terms, it is not possible to
close $[\delta_1,\delta_2] (X^I)_A^+$ onto translations (up to
gauge transformations). A similar result is also true for the
$T^A_-$ components.


Working in the gauge demanded by the invariance of the action
(\emph{cf.} discussions of the opening of section 4.3),%
\be\label{susy-gauge}%
v^\mu {\tilde A}_{\mu AD}=\tilde{\Lambda}_{AD},%
\ee%
and we remain with%
\be\label{susy-closing-XI-component-2}%
\begin{split}%
[\delta_1,\delta_2] (X^I)^+_D &=v^\mu \partial_\mu (X^I)^+_D+
\chi_{AD}^+\ (X^I)^-_A\cr%
[\delta_1,\delta_2] (X^I)^-_D &=v^\mu \partial_\mu (X^I)^-_D-
\chi_{AD}^-\ (X^I)^+_A.
\end{split}%
\ee%
That is, the supersymmetry will close only if $\chi^\pm_{AD}$ are
vanishing (on the ``physical Fock space of the theory''). Recalling
that, with the complex valued $(X^I)^\pm_A$ we have introduced twice
as much fields, there is the possibility of closing the
supersymmetry on the physical Fock space which only involves a
specific half of the degrees of freedom. As we show there is indeed
such a possibility.

$\bullet$ Closure of \susy\ on fermions, after using the equation
of motion of fermions, leads to \cite{BL2}%
\be\label{susy-closing-Psi}%
[\delta_1, \delta_2]\Psi=v^\mu D_\mu \Psi - V_{JK}\ [\Psi,
X^J, X^K, T]\ . %
\ee%
The same analysis  presented for $X^I$'s also holds for fermions
and in \eqref{susy-gauge} gauge the above reduces to
\eqref{susy-closing-XI-component-2} with $(X^I)^\pm_A$ replaced
with $\Psi^\pm_A$. Therefore, closure of \susy\ for fermions
demands a similar condition as the scalars, the point to be
discussed momentarily.

$\bullet$ The closure of \susy\ for the gauge fields is more
involved. Performing the analysis, we find that in $[\delta_1,
\delta_2]\tilde A_{\mu AB}$ there is a term
proportional to (see eq.(35) of \cite{BL2})%
\be\label{N=8SUSY-cond}%
-\frac{i}{3}\left(\bar\epsilon_2 \gamma_\mu
\Gamma^{IJKL}\epsilon_1\right)\ Tr (X^I[[X^J,X^K,X^L, T],T^A_+,
T^B_-, T])\ . %
\ee%
This term vanishes for any two arbitrary $3d$ fermions $\epsilon_1,
\epsilon_2$ and any choice of $A, B$ indices, once we recall the
extended \FI\ \eqref{FI-extended}, \eqref{push-thru-trace} and the
totally antisymmetry of $\Gamma^{IJKL}$. Following the computations
of \cite{BL2} and using the equation of motion of the gauge field we
obtain%
\be\label{susy-closing-gaugefield}%
[\delta_1, \delta_2]{\tilde A}_{\mu AB}=v^\nu {\tilde F}_{\mu\nu
AB} -
D_\mu \tilde\Lambda_{AB} , %
\ee%
where%
\[
\tilde F_{\mu\nu AB}= \partial_{\mu} {\tilde A}_{\nu
AB}-\partial_{\nu} {\tilde A}_{\mu AB}+ {\tilde A}_{\mu AC} {\tilde
A}_{\nu CB}-{\tilde A}_{\nu AC}{\tilde A}_{\mu CB} .
\]
In the gauge \eqref{susy-gauge}, we see that $[\delta_1,
\delta_2]\tilde A_\mu$,  closes on translations without any extra
$\chi^\pm$-type terms.

\subsubsection{Projection onto the supersymmetric Hilbert space}

Although the \susy\ transformations are compatible with parity (see
appendix C), $X^I$ are not parity invariant and hence  the Fock
space constructed from operators built upon $X^I$ is not parity
invariant. One may hope that the above \susy\ non-closure will be
resolved on the ``parity invariant'' sector of the Fock space. As
can be seen from the closure analysis of previous subsections the
supersymmetry closure implies  $\chi^\pm_{AB}=0$, which obviously
cannot be realized while keeping the $SO(8)$ invariance of the Fock
space. We are hence forced to compromise the $SO(8)$ covariance of
the states.\footnote{To render the action invariant, there is one
other option: To restrict the theory to specific (BPS)
configurations over which $\partial_\mu J^\mu$ vanishes. These
specific configurations should, however, form a \emph{closed sector}
in the Hilbert space. We will briefly explore this possibility in
section 6.}

The $\chi^\pm_{AB}=0$ condition can, however, be met on a smaller
set of states and fermionic (supersymmetry) transformations. It
turns out that the largest sector in the Hilbert space of the
theory for which $\chi^\pm_{AB}$ vanishes is the part which is
invariant under $SO(6)\times U(1)\simeq SU(4)\times U(1)\in
SO(8)$. To see this we should perform a specific ``projection''
onto this $SU(4)\times U(1)$ invariant sector. Let us start with
the $(X^I)^\pm$. Instead of a generic function (operator made) of
eight complex valued $(X^I)^\pm$ we
project onto the functions (states) made out of four complex scalars%
\be\label{su(4)-scalar}%
Z^\alpha=X^\alpha_++iX^{\alpha+4}_+,\qquad {\bar
Z}_{\alpha}=(Z^{\alpha})^*=X^\alpha_--iX^{\alpha+4}_-,\qquad
\alpha=1,2,3,4\ . %
\ee%
It is evident that $Z^\alpha$ and ${\bar Z}_\alpha$ transform as
$\mathbf{4}$ and $\mathbf{\bar 4}$ of $SU(4)$ and under the $U(1)$
$Z_\alpha\to e^{i\xi} Z_\alpha$. To distinguish this $U(1)$ symmetry
from the one introduced in \eqref{global-U(1)b} we denote it by
$U(1)_\xi$. That is, e.g. $Z_{\alpha}$ is in $\mathbf{4}_{+1}$ and
${\bar Z}_\alpha$ in $\mathbf{\bar 4}_{-1}$ of $SU(4)\times
U(1)_\xi$.

As discussed in the end of section 4.2 our $(X^I)^\pm$ fields are
also charged under the global $U(1)_\lambda$. It is evident that
$Z_\alpha$ carry charge $+1$ and ${\bar Z}_\alpha$ charge $-1$ of
the $U(1)_\lambda$; that is, $Z_\alpha$ is in $(+1,+1)$ and ${\bar
Z}_{\alpha}$ in $(-1,-1)$ representation of $U(1)_\lambda\times
U(1)_\xi$. We comment that
\[%
(Z^\alpha)_{parity}=X^\alpha_-+iX^{\alpha+4}_-\neq {\bar Z}_{\alpha}%
\]%
and as such  under parity the $SU(4)$ and $U(1)_\xi$ representation
remains intact while the $U(1)_\lambda$ charge changes sign.
$(Z^\alpha)_{parity}$ and $({\bar Z}_\alpha)_{parity}$ are hence
respectively in $(-1,+1)$ and $(+1,-1)$ representation of
$U(1)_\lambda\times U(1)_\xi$. Restricting to the combination of
$X^I$'s which are made out of $Z_\alpha$ and ${\bar Z}^\alpha$ then
means that we project onto states made out of  linear combination of
$(X^I)^\pm$ fields for which the product of their
$U(1)_\lambda\times U(1)_\xi$ is positive. In this way half of the
degrees of freedom of $X^I$'s are projected out.
 We perform a similar decomposition for
the complex valued fermionic fields $\Psi^\pm$ which are in
$\mathbf{8_s}$ of $SO(8)$ and decompose them into
$\mathbf{4}_{+1}+\mathbf{\bar 4}_{-1}$ of $SU(4)\times U(1)_\xi$
fermions and work with the states made out of linear combinations of
$\Psi$'s the product of their $U(1)_\lambda\times U(1)_\xi$ charges
is $+1$.

The \susy\ variation parameters $\epsilon$ do not carry $\pm$
indices (they are neutral under $U(1)_\lambda$) and are in
 $\mathbf{8}_c$ of $SO(8)$, as well as being a $3d$ anti-Majorana fermion.
The $\mathbf{8}_c$ decomposes to $\mathbf{6}_0+
\mathbf{1}_{-2}+\mathbf{1}_{+2}$ of $SU(4)\times U(1)_\xi$. If
together with working with the configurations (states) which are
made out of $Z_\alpha$ and its fermionic counterpart, we
\emph{restrict} ourselves to the \susy\ transformations generated by
$\epsilon$ which are in $\mathbf{6}_0$, $\chi$-terms vanish.
To see this let us consider $\chi^+_{AB}$. Vanishing of
$\bar\epsilon_2\chi^+_{AB}\epsilon_1$ may be seen recalling the form
of $\chi_{AB}^\pm$ \eqref{chipm-def} and noting that the $V_{JK}$
part is in $(\mathbf{6}\times \mathbf{6})_{A.S.}=\mathbf{15}$ while
the $X^J X^K$ piece is in $(\mathbf{4}_{+1}\times
\mathbf{4}_{+1})_{A.S.}=\mathbf{6}_{+2}$. Since $\mathbf{15}\times
\mathbf{6}$ does not give a singlet of $SU(4)$,
$\bar\epsilon_2\chi^+\epsilon_1$ vanishes. Similarly one can argue
that $\chi^-$ vanishes. In this way out of 16 independent fermionic
transformations  only 12 of them close onto the \susy\ algebra.

To summarize, restricting the fields to $Z_{\alpha}$ and their
fermionic counterpart the \susy\ transformations  which are
generated by $\epsilon$'s in $\mathbf{6}_0$ of $SU(4)\times
U(1)_\xi$ close and our gauge invariant action will describe  a
theory which has $3d$, ${\cal N}=6$ supersymmetry.


Although for a generic configuration we are dealing with an \Ns
theory, there are still large class of states (configurations)
which exhibit more fermionic symmetries than expected from the \Ns
theory. Let us consider states of the form ${\cal O}^{I_1\cdots
I_l}= Tr(X^{I_1}X^{I_2}\cdots X^{I_{l}})$ where the trace is over
the $2N\times 2N$ matrices.\footnote{Recalling that
$(\sigma^+)^2=(\sigma^-)^2=0$ and that $\sigma^\pm$ are traceless
for odd $l$ ${\cal O}^{I_1\cdots I_l}$ vanishes and for even $l$%
\[%
 {\cal O}^{I_1\cdots I_l}=Tr_N\left((X^{I_1}_+ X^{I_2}_-
X^{I_3}_+ X^{I_4}_-\cdots X^{I_l}_-)+(X^{I_1}_- X^{I_2}_+ X^{I_3}_-
X^{I_4}_+\cdots X^{I_l}_+)\right),\]%
where $Tr_N$ is over $N\times N$
matrices. Gauge invariant operators which are constructed out of
trace over $2N\times 2N$ matrices are neutral under the
$U(1)_\lambda$. Moreover, ${\cal O}_{I_1\cdots I_l}$ type operators
are also parity invariant.} It is a straightforward computation to
show that under two successive \susy\ transformations
$[\delta_1,\delta_2]{\cal O}_{I_1\cdots I_l}=v^\mu\partial_\mu{\cal
O}_{I_1\cdots I_l}$. One can repeat the same computation with
operators in which some of the $X^I$'s are replaced with $SO(8)$
fermions $\Psi$. For these operators, too, two successive \susy\
transformations close onto the derivative of the operator. For the
operators which involve covariant derivative of $X^I$ or $\Psi$,
e.g. $Tr(X^ID_\mu X^J)$, the \susy\ does not close onto
translations; for these operators there remain some terms stemming
from the $\chi^\pm$ terms in \eqref{susy-closing-XI-component-2}. We
note that the set of ${\cal O}_{I_1\cdots I_l}$ type operators
include the chiral primaries. Therefore, although in general our
theory enjoys \Ns \susy, there are large classes  of gauge invariant
BPS states which can preserve more fermionic symmetries than the
ones expected from an \Ns theory. In section 6 we will discuss
examples of such BPS states.

We point out that if we rewrite the action implementing the
restriction of the fields to $\mathbf{4}_{+1}$ and $\mathbf{{\bar
4}}_{-1}$ our theory reduces to the representation of the ABJM model
in terms of (non-totally antisymmetric) three-algebras
\cite{BL3-bracket-N6}. The structure constants of their model is
hence equal to our $f^{ABCD}$. In this construction the $T^A_\pm$
are not appearing explicitly and one only deals with $N\times N$
matrices.

Before closing this section we stress that as discussed in section 3
the $N=2$ case is special in the sense that the $f^{abcd}$
coefficients \eqref{f-explicit} vanish. As shown in the appendix B,
our $u(2)$-based extended three-algebra is a double copy of the
$so(4)$-based Bagger-Lambert three-algebra. One can use this
observation to project out half of the excessive degrees of freedom
of the $u(2)$ theory. Projecting onto  the $\Phi^+_a=\Phi^-_a,\
\Phi^+_0=-\Phi^-_0$ sector (where $\Phi$ is $X^I$ or $\Psi$) and
$a=1,2,3$, our theory reduces to the Bagger-Lambert theory. This
projection explicitly keeps the $SO(8)$ invariance as well as \susy.
(After this projection one may explicitly check that for this case
there is a gauge, the one worked out in \cite{BL2}, in which the
action becomes invariant under 16 \susy\ transformations.) We
emphasize that this is a different projection than the $SU(4)\times
U(1)$ invariant one used to earlier. In this sense our analysis
shows how the Bagger-Lambert and ABJM theories for $N=2$ are
different.

\section{The $su(N)\times su(N)$ Chern-Simons  representation}

As argued the closure of the extended three-algebra, with the
working assumption that $t^A$ are generators of a (semi)-simple Lie
algebra, fixes the Lie-algebra to be $u(N)$ in its $N\times N$
representation. Here we rewrite the theory using the explicit
representation of $f^{ABCD}$ in terms of $su(N)$ $f$ and $d$ tensors
and remove the four-brackets. Let us start with the gauge fields
$A_{\mu AB}$ and ${\tilde A}_{\mu AB}$. Using
\eqref{f-explicit}, \eqref{tildeA} can be written as%
\be\label{tildeA-uN}%
\begin{split}%
{\tilde A}_{\mu cd} &= f_{cde} A_{\mu e}+i d_{cde} B_{\mu e},\cr
{\tilde A}_{\mu a0} &={\tilde A}_{\mu 0a}=\frac{2i}{\sqrt{2N}}\
B_{\mu a}, \cr%
{\tilde A}_{\mu 00}&=0\ ,\qquad \sum_a {\tilde A}_{\mu aa}=\sum_A
{\tilde A}_{\mu AA}=0
\end{split}%
\ee%
where%
\be\label{A-B-gauge-fields}%
\begin{split}%
A_{\mu e} &\equiv -\frac{i}{12}\left( d_{abe} A_{\mu
ab}+\frac{2}{\sqrt{2N}}(A_{\mu 0e}+A_{\mu e0})\right)\ , \cr %
B_{\mu e} &\equiv -\frac{1}{12}f_{abe} A_{\mu ab},
\end{split}
\ee%
are two \emph{real $su(N)$ valued} gauge fields. The reality of
$A_{\mu a}$ and $B_{\mu a}$ gauge fields is a result of
\eqref{A-hermiticity}.

The covariant derivative of the matter fields $\Phi$ in terms of
these $su(N)$ gauge fields take the form%
\be\label{cov-der-suN}%
\begin{split}%
(D_\mu \Phi)^+_d &=\partial_\mu \Phi^+_d- (f_{cde} A_{\mu e}+ i
d_{cd e} B_{\mu e})\ \Phi^+_c-\frac{2i}{\sqrt{2N}} B_{\mu d}
\Phi^+_0\ ,\cr%
 (D_\mu \Phi)^+_0 &=\partial_\mu \Phi^+_0-\frac{2i}{\sqrt{2N}} B_{\mu c}
\Phi^+_c\ .%
\end{split}%
\ee%
Note that $(D_\mu\Phi)^-_A=\left((D_\mu\Phi)^+_A\right)^*$.

Recalling the behavior of the gauge field under parity
\eqref{parity-gauge-field}, we learn that  under parity $A_{\mu a}$
behaves as a vector while $B_{\mu a}$ transforms as a pseudovector.
Rewriting the twisted \cs part of the action in
terms of $A$ and $B$ gauge fields we find%
\be\label{twisted-cs}%
{\cal L}_{\cs}= \frac{1}{2}\epsilon_{\mu\nu\alpha}\left[-12 B_{\mu
a}\partial_{\nu} A_{\alpha a}+2 f_{abc} (B_{\mu a}B_{\nu b}
B_{\alpha c}+3 B_{\mu a} A_{\nu b} A_{\alpha c})\right]\ .%
\ee%
With a vector $A_{\mu a}$ and pseudovector $B_{\mu a}$ it is clear
that the above action is parity invariant.

Upon the field redefinition%
\be\label{GL-gauge-fields}%
R_{\mu a}=A_{\mu a}- B_{\mu a},\qquad L_{\mu a}=
A_{\mu a}+ B_{\mu a}\ , %
\ee%
\eqref{twisted-cs} takes the form%
\be\label{GL-cs}%
{\cal L}_{\cs}={\cal L}_{\cs\ R }-{\cal L}_{\cs\ L}\ , %
\ee%
where%
\be\label{G-cs}%
{\cal L}_{\cs\ R }=\frac{3}{2 }\ \epsilon_{\mu\nu\alpha}\left[
R_{\mu a}\partial_{\nu} R_{\alpha a}-\frac{1}{3} f_{abc} R_{\mu
a}R_{\nu
b} R_{\alpha c}\right] ,%
\ee%
and similarly for ${\cal L}_{\cs\ L}$. Therefore, the \cs part of
the action \eqref{BLG-action-four-bracket} is nothing but the
standard $su(N)\times su(N)$ \cs action. The level of the \cs of the
two $su(N)$ \cs factors are equal but with the opposite sign and the
parity exchanges the two $su(N)$ factors. Under the $U(1)_\lambda$,
$A_{\mu AB}$ remains invariant (\emph{cf.} \eqref{global-U(1)b}) and
as a result the $su(N)$ gauge fields $R_{\alpha}$ and $L_\alpha$
also remain invariant.

In usual conventions for the  \cs theories, we have obtained a \cs
theory at level $12\pi$. There is the possibility of getting the \cs
theory in an arbitrary level $k\in\mathbb{Z}$. In order this we may
keep the form of the action we start from
\eqref{BLG-action-four-bracket}, and similar to \cite{M2onOrbifold},
replace the structure constants $f^{ABCD}$ by $f^{ABCD}/12\pi k$
(this scaling does not change the \FI\ and closure conditions). This
may be achieved by changing the normalization of the $u(N)$
generators $t^A$ to $t^A/\sqrt{12\pi k}$.

Starting from (\ref{gauge-sym}c), after appropriate decomposition
of the gauge transformation parameter $\tilde \Lambda_{AB}$ and
using the $su(N)$ identities listed in the appendix A, one can
work out the behavior of the $R_\mu$ and $L_\mu$ gauge fields under gauge transformation%
\be\label{GL-gauge-trans}%
\begin{split}%
\delta_{gauge} R_{\mu a} &= \partial_\mu \rho_a-f_{abc} R_{\mu b}
\rho_c, \\
\delta_{gauge} L_{\mu a} &= \partial_\mu \lambda_a-f_{abc} L_{\mu
b} \lambda_c,
\end{split}%
\ee%
which as expected are two $su(N)$ gauge transformations.

Now let us study behavior of the matter fields under the above
$su(N)\times su(N)$ factors. From (\ref{gauge-sym}b) and after
straightforward, but lengthy algebra using $su(N)$
identities listed in the appendix A, we find that%
\be\label{matter-gauge-trans-suN}%
\begin{split}%
\delta_{gauge} \Phi^+ &= i[\chi^1,\Phi^+]+i\{\chi^2, \Phi^+\}\cr
\delta_{gauge} \Phi^- &= i[\chi^1,\Phi^-]-i\{\chi^2, \Phi^-\}\ ,
\end{split}%
\ee%
where $\Phi^\pm$ includes both the $su(N)$ and $u(1)$ components,
respectively $\Phi^\pm_a$ and $\Phi^\pm_0$, of the fields and
\be%
\chi^1_a=\half(\rho_a-\lambda_a)\ ,\qquad \chi^2_a=\half(\rho_a+\lambda_a)\ .%
\ee%
Note that $\chi^i$, like $\lambda$ and $\rho$, are $su(N)$ (and
not $u(N)$) valued. From \eqref{matter-gauge-trans-suN} one can
read the form of the finite gauge transformations of $\Phi^\pm$:
\be\label{matter-gauge-trans-suN-finite}%
\begin{split}%
\Phi^+ \longrightarrow \ {\tilde \Phi}^+=e^{i\lambda}\ \Phi^+\
e^{-i\rho}\ , \qquad \Phi^- \longrightarrow \ {\tilde
\Phi}^-=e^{i\rho}\ \Phi^-\ e^{-i\lambda}\ .
\end{split}%
\ee%
That is, $\Phi^\pm$ are in the bi-fundamental representation of
$su(N)\times su(N)$. As discussed under the (global) $U(1)_\lambda$
$\Phi^\pm$ carry charge $\pm 1$.

For completeness we also present the explicit form of the fermionic
(supersymmetry) transformations in terms of the \cs fields. The
scalars and fermions have basically the same form as given in
(\ref{SUSY-trans-four-bracket}a,b) and for the gauge fields
(\ref{SUSY-trans-four-bracket}c) becomes%
\be\label{susy-cs}%
\begin{split}%
\delta_{susy} A_{\mu }&=\frac{1}{12}\bar\epsilon \gamma_\mu
\Gamma^I \bigl(\{(X^I)^+, \Psi^-\}-\{(X^I)^-, \Psi^+\}\bigr)\cr %
\delta_{susy} B_{\mu }&=-\frac{i}{12}\bar\epsilon \gamma_\mu
\Gamma^I \bigl([(X^I)^+, \Psi^-]+[(X^I)^-, \Psi^+]\bigr)\ .
\end{split}
\ee%
It should be noted that the above will become \susy\ transformations
once the projection to $SU(4)\times U(1)$ sector is performed.

 So far we have presented our model in terms of an $su(N)\times
su(N)$ Chern-Simons theory with explicit $SO(8)\times U(1)_\lambda$
symmetry. As argued in previous section out of the 16 independent
fermionic variations introduced in \eqref{SUSY-trans-four-bracket}
only 12 of them can lead to symmetries of the action. The \susy\
closes only on a $SU(4)\times U(1)_\xi\times U(1)_\lambda$ invariant
sector of the physical Fock space of the theory. Once the theory is
rewritten in terms of the fields over which the supersymmetry closes
our theory becomes the $3d$, \Ns $su(N)\times su(N)$ \cs\ theory.
Our theory is hence closely related to the ABJM model. \footnote{We
should, however, note that as discussed earlier the $Z_\alpha$ and
${\bar Z}_\alpha$ are not related by worldvolume parity; they are
related by a product of parity and $U(1)_\xi$ charge conjugation. In
this sense the ABJM theory, even for $N=2$ is different than the
Bagger-Lambert theory.}

The model ABJM proposed to describe the low energy dynamics of $N$
M2-branes (on $\mathbf{C}^4/\mathbb{Z}_k$ orbifold) is, however, a
$u(N)\times u(N)$ theory (rather than \sun2). This model is related
to our model upon gauging two extra global $U(1)$'s. One of them is
the $U(1)_\lambda$ and the other is the ``center of mass'' $U(1)$,
$U(1)_{cm}$. Recalling that $\Phi^\pm$ fields are in the
bi-fundamental of the \sun2 
\eqref{matter-gauge-trans-suN-finite} one may simply gauge the
$U(1)_{cm}$ symmetry without the need to add any additional
interactions for $\Phi$'s, once we identify the $U(1)_{cm}$ with the
diagonal part of the $u(1)$'s in $u(N)\times u(N)$. Gauging
$U(1)_{cm}$, then only amounts to adding the corresponding $U(1)$
\cs\ term. As discussed in section 4.3.2 the $U(1)_\lambda$ charge
changes sign under parity while the $U(1)_{cm}$ charge remains
invariant. This is compatible with identifying $U(1)_{cm}$ with the
diagonal $U(1)$ and $U(1)_\lambda$ is the anti-symmetric
combinations of the two $U(1)$'s in $U(N)\times U(N)$. \footnote{As
argued in \cite{ABJM} the $3d$ Chern-Simons $U(1)$ gauge theory has
the peculiar feature that its equation of motion is $^*F=J$ ($J$ is
the $U(1)$ currents) and hence we have a global symmetry generated
by the conserved current $J = ^*F$. The $U(1)_b$ symmetry in the
ABJM model, which is a part of the R-symmetry of the M2-brane
theory, is the global $U(1)$ generated by the diagonal $U(1)$ part
of the $U(N)\times U(N)$ gauge symmetry (the $U(1)_\xi$ in our
notation) through $J_d = ^*F_d$. We thank Ofer Aharony for
clarifying comment on this point.} In the theory in which
$U(1)_{cm}$ is gauged, even after fixing the gauge, we remain with a
$\mathbb{Z}_k$ part of the $U(1)$ and hence the $Z_\alpha$ are
defined up to $\mathbb{Z}_k$ rotations. Therefore,  this theory
describes M2-branes on $\mathbf{C}^4/\mathbb{Z}_k$ orbifold. As
discussed in \cite{ABJM}, just gauging the two extra $U(1)$'s does
not bring our \sun2 theory to the ABJM model and one should consider
two points: $U(N)\simeq (SU(N)\times U(1))/Z_N$ and that in the
$SU(N)\times U(1)$ theory, despite the fact that in general the \cs
levels for the $U(1)$ and $SU(N)$ parts could be different, in the
$U(N)$ theory they are taken to be equal.

After relating our theory to the ABJM model, their arguments for the
physical states also apply to ours. Physical states of our theory
can be those which are invariant under $U(1)_\lambda$. In the
language of our three-algebra representation, these states could be
constructed by taking trace over $2N\times 2N$ matrices, like the
${\cal O}^{I_1I_2\cdots I_l}$ operators of last section.
\footnote{It is instructive to note that the bi-fundamental nature
of the ABJM fields $Z_\alpha$, dictating  that the gauge invariant
combinations should involve $Z_\alpha {\bar Z}_\beta$ or ${\bar
Z}_{\alpha}Z_\beta$ which fall into adjoint representations of
either of the $U(N)$ factors, is naturally encoded in  our $2N\times
2N$ matrices. This is because of \eqref{Tpm2=0} which implies that
$X^IX^J=(X^I)^+(X^J)^-+(X^J)^-(X^I)^+$.} As discussed in
\cite{ABJM}, there are also states which carry $k$ units of the
$U(1)_\lambda$ charge, those which have particular Wilson lines
attached.

 We note that after gauging the two $U(1)$'s the theory cannot be
expressed in terms of the (extended) three-algebra anymore.

\section{Relation to the theory of M2-branes}

The \3dNe (or its \Ns version) SCFT should arise as the low energy
effective field theory limit of coincident multi M2-branes on flat
space (or its orbifold). Here we argue that  the action
\eqref{BLG-action-four-bracket} for our $u(N)$-based extended
three-algebra and after restricting (``projecting'') to $SU(4)$
invariant sector of the Hilbert space, describes theory of $N$
M2-branes. In addition we bring arguments clarifying the need for
the projection.

\subsection{Pair-wise M2-brane picture}

It is well known and understood that when $N$ D-branes of string
theories sit on top of each other we see the structure of a $u(N)$
gauge theory \cite{Witten}. For the special case of D3-branes this
theory (in the low energy limit) is the $u(N)$ $4d$ SCFT. The
enhancement of the gauge symmetry to $u(N)$ in the D-brane case is
facilitated by the (perturbative) description of D-branes in terms
of open strings ending on or stretched between D-branes. In the
coincident limit the lowest modes of these open strings become
massless and hence cause the gauge symmetry enhancement (inverse
of Higgs mechanism). The above picture for D-branes and open
strings stretched between them is valid for any  pair of D-branes
in a system of $N$ D-branes \cite{Witten}.

\begin{figure}[t]
\begin{center}
\includegraphics[scale=0.6]{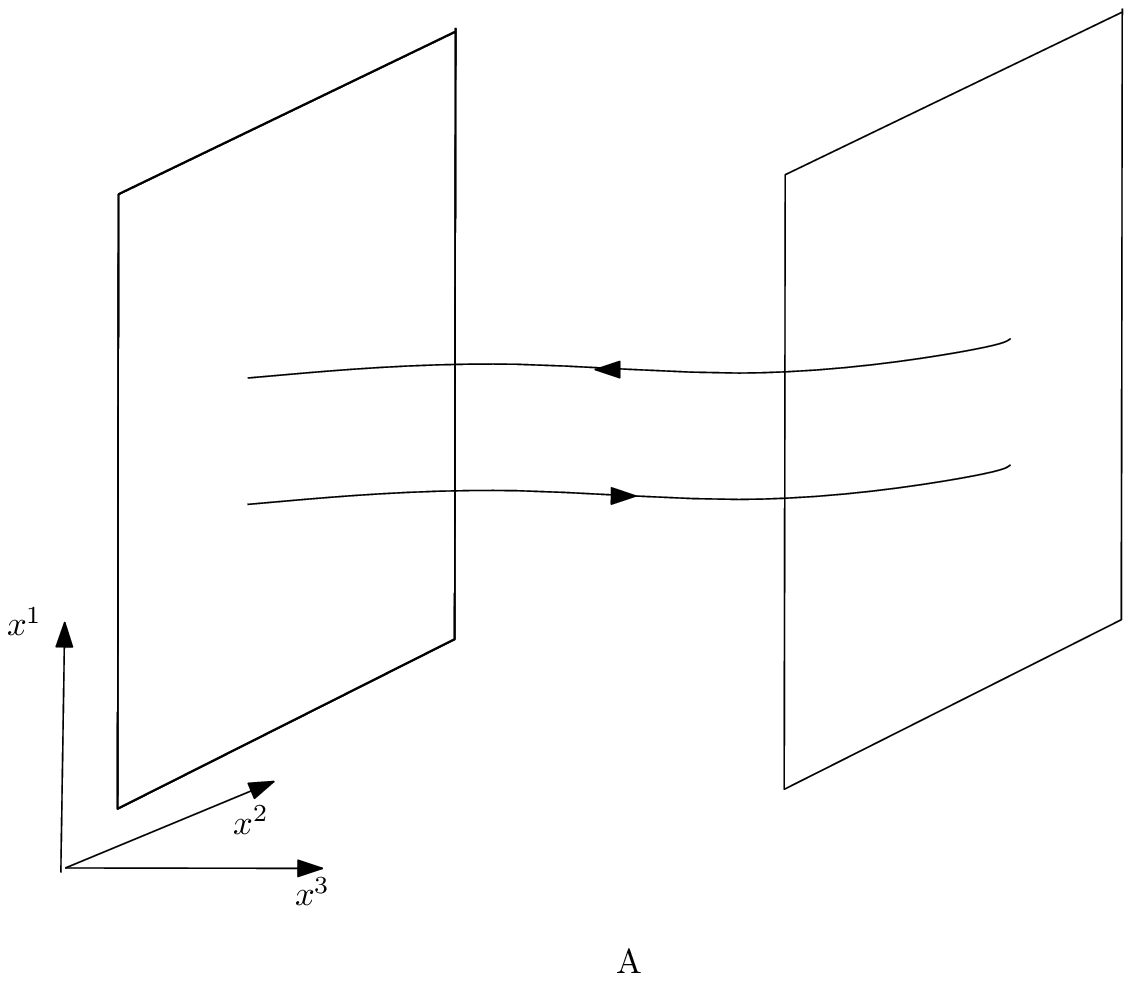}\includegraphics[scale=0.6]{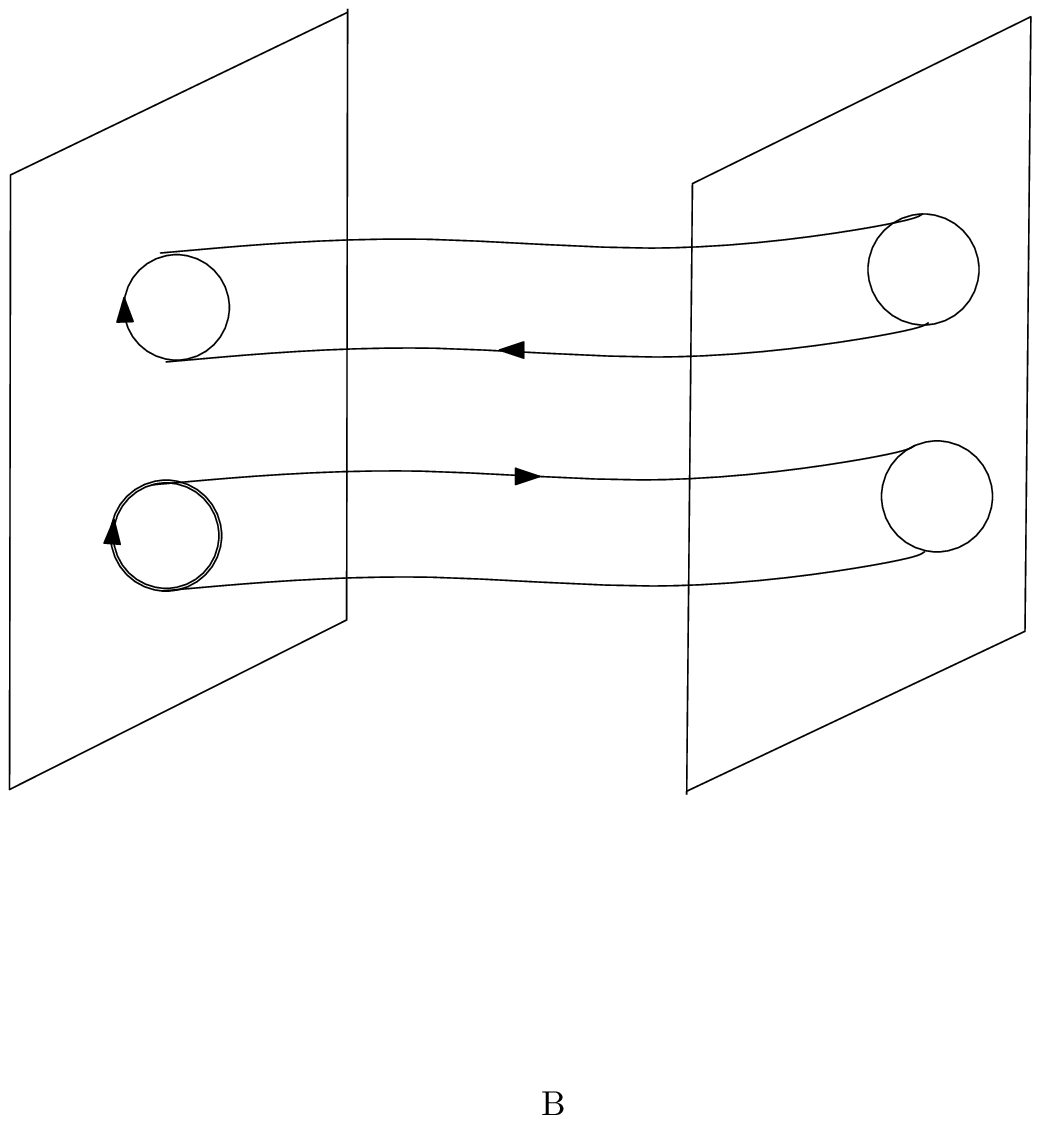}\caption{
The figure on the left shows the open string anti-open string pair
stretched between two parallel $D2$-branes along $012$ directions
while separated in $x^3$. The figure on the right shows the same
system after uplifting to M-theory, where the open string pair now
appear as open membrane anti-open membrane pair. Note that the
open membranes wrap the $11^{th}$ circle in the same orientation.}
\label{Fig1}
\end{center}
\end{figure}

\begin{figure}[ht]
\begin{center}
\includegraphics[scale=0.5]{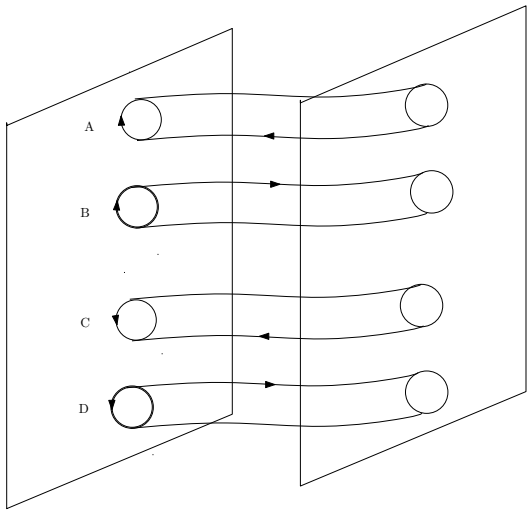}\caption{
There are two options for open membrane anti-open membrane pairs
stretched between two M2-branes. If we assume the M2-branes to be
along $012$ directions, these open membranes are along say $034$
($x^3$ is the direction the M2-branes are separated and $x^4$ is
along the circular part of the open membranes). The open membrane
pair A and B are mapped to C and D under worldvolume parity. Note
that the circular direction on the open membranes is just for
illustrative purposes and in terms of our matrices this part is
associated with the $\sigma^\pm$ parts. In terms of what we have in
the Figure, i.e. A and D are associated with $T^A_+$ and B and C
membranes with $T^A_-$. Note that as far as the M2-brane charge is
concerned the $A$ and $D$ and, $B$ and $C$ open membranes are
indistinguishable and hence we need to mod out the ``excess of
degrees of freedom'' we have introduced in our setting.}
\label{Fig2}
\end{center}
\end{figure}

The above ``pair-wise'' picture does not readily generalize to the
M2-branes, as here we do not have  the open strings picture.
Nonetheless, we have open membranes stretched between two
M2-branes. To see how these open membranes come about, let us
start with two parallel D-branes in $10d$ IIA string theory. As
shown in Fig.\ref{Fig1}A there are (virtual) open string
anti-string pairs stretched between the D2-branes. These open
strings are oriented and the difference between the open string
and anti-open string is the orientation; they are related by the
worldsheet parity. When uplifted to M-theory the D2-branes become
M2-branes while the stretched open strings become open membranes
and anti-open membranes (see Fig.\ref{Fig1}B).
Had we directly started in the $11d$ M-theory, as membrane
worldvolume have two spatial directions, unlike the string case and
as depicted in Fig.\ref{Fig2}, there are two distinct options for
open membrane anti-open membrane pair. These two pairs are related
by the worldvolume parity. On the other hand, from the M2-brane
viewpoint not all the four possibilities in the Fig.\ref{Fig2} are
independent, explicitly, $A$ and $D$ open membranes and $B$ and $C$
open membranes cannot be distinguished by their M2-brane charge.

In the same spirit as D-branes, for the case of $N$ M2-branes, we
expect that we should be dealing with $2N\times 2N$ matrices. In our
realization the $2\times 2$ $\sigma^\pm$ part of the $T^A_\pm$
generators basically account for this ``doubling'' of the degrees of
freedom corresponding to the open membrane pairs (compared to the
open string case). However, as discussed not all the degrees of
freedom of these stretched membranes are physically independent and
moreover, not all of them can appear in the supersymmetric Fock
space of the M2-brane theory; we need to mod out half of them.
Restricting to the sector over which the \susy\ transformations
close (onto the $3d$, \Ns\!\!) these extra degrees of freedom are
removed. This sector is identified with part of the Fock space, {the
physical Fock space}, which is made out of functions of combinations
of $X$'s and $\Psi$'s which the $U(1)_\lambda$ and $U(1)_\xi$ have
the same sign. In addition, this picture also sheds light on the
\sun2 structure.

 Starting from this
M2-brane picture, compactifying down to $10d$ IIA theory, however,
only one of the two open membrane pairs  survive the \susy\
requirement. Supersymmetry demands that open membrane pairs should
have the same orientation on the $11^{th}$ circle (in Fig
\ref{Fig2}, i.e. A and B or C and D pair). Therefore, at the IIA
and D2-brane level we only see a single $su(N)$ factor.
\footnote{We should stress that, since we do not have the spectrum
of open membranes, unlike the case of strings, our open membrane
picture should be only taken as a helpful and suggestive pictorial
way of presenting the $su(N)\times su(N)$ structure.}

\subsection{Analysis of BPS  states}

In the previous subsection, based on the stretched open membrane
picture,  we argued that we expect an \sun2 \cs theory (of course
plus the gauging of the two extra $u(1)$'s) to describe $N$
M2-branes (on an orbifold). To substantiate this result we analyze
the BPS states of our theory.

Recalling that not all the generic configurations of our $X^I$ and
$\Psi$ fields close the \susy\ ``algebra'' resulting from the
fermionic transformations \eqref{SUSY-trans-four-bracket}. As
discussed all the (bosonic) configurations which are formed out of
$Z_\alpha$ fall into representations of \Ns algebra. However, there
could be some states preserving more \susy\ than expected from the
\Ns theory. In order not to lose the extra supersymmetry of these
states, we perform the BPS analysis as follows. First we find
solutions to $\delta_{susy} \Upsilon=0,$ with $\Upsilon$ being
either of $X^I, \Psi,$ or $A_{\mu AB}$ fields and ignoring the fact
that  not all the configurations which satisfy $\delta \Upsilon=0$
are necessarily falling into the representations of \Ne or \Ns
superPoincar\'e algebra. As the second step we check whether these
particular (BPS) configurations/states indeed satisfy the closure of
\susy\ algebra. In order this we check if $[\delta_1,\delta_2]$,
with $\delta$ given in \eqref{SUSY-trans-four-bracket}, on the
specific configuration in question is equal to $v^\mu\partial_\mu$
on that configuration.

\subsubsection{Half-BPS states}

As the candidate for $N$ M2-branes on the $11d$ flat space (or its
orbifold) the moduli space of 1/2 BPS configurations of our model
must be $R^{8N}/S_N$ (or ($\mathbf{C}^4/\mathbf{Z}_k)^N/S_N$). The
half BPS sector of our model is the one for which the
right-hand-side of all supersymmetry variations
\eqref{SUSY-trans-four-bracket} vanishes for any arbitrary
fermionic transformation parameter $\epsilon$. Variations of the
bosonic fields identically vanish for a pure bosonic
configuration. Variation  of fermions vanish for arbitrary
$\epsilon$ only when
the two terms in $\delta\Psi$ vanish independently, i.e.%
\bse\label{1/2BPS}%
\begin{align}%
D_\mu X^I &=0 \ ,\\
[X^I, X^J, X^K, T]&= 0.%
\end{align}%
\ese%

When (\ref{1/2BPS}a) holds and the fermionic fields are turned
off, the equations of motion for the two $su(N)$ gauge fields
imply that both the gauge fields have flat connection and hence
they can be set to zero in appropriate gauge. In this gauge,
(\ref{1/2BPS}a) implies $\partial_\mu X^I=0$. (\ref{1/2BPS}b) is
satisfied if and only if%
\be\label{XXX}%
[(X^I)^+, (X^J)^-]=0\ ,\qquad [(X^I)^+, (X^J)^+]=0\ ,%
\ee%
where $(X^I)^\pm$ are the $N\times N$ matrices and may be defined
through taking trace over $2\times 2$ parts of the $2N\times 2N$
matrices, explicitly: $(X^I)^\pm=Tr_{2\times 2} \left( X^I\cdot
(\id_N\otimes \sigma^\mp)\right).$ \eqref{XXX} is satisfied for any
diagonal $N\times N$ matrices (on the elements on the diagonal
complex valued). To find the moduli space of physical solutions,
however, we still need to restrict ourselves to the \Ns
supersymmetric sector. This is done by restricting to diagonal
$Z_\alpha$ matrices. This removes half of the solutions, rendering
the solutions to $8N$ real parameters. The analysis then becomes
identical to that of ABJM \cite{ABJM} with a minor difference on the
number of conserved supercharges: Recalling \eqref{susy-closing-XI}
and \eqref{1/2BPS}, it is readily seen that $[\delta_1,\delta_2]$
over these configurations vanish. Moreover, for these configurations
and also  the other states which fall into the same \Ne
supermultiplet the variation of the action \eqref{susy-action-var}
vanishes. Therefore, these configurations form a sector which is
invariant under all the 16 ``supersymmetry'' variations are 1/2 BPS
in the sense of \Ne.



\subsubsection{1/4-BPS, Basu-Harvey configuration}

There are much further options for less BPS cases. Here we consider
the 1/4 BPS state which corresponds to M2-brane along $056$ ending
on an M5-brane along $012345$, the Basu-Harvey configuration
\cite{BH}. \footnote{For the analysis of finding M5-M2 solutions in
the ABJM model see \cite{Terashima}. Analysis of some other BPS or
time-dependent non-BPS configurations of the BLG or ABJM models may
be found in \cite{Other-solutions}.} Turning off the fermions, the
BPS configurations are obtained as solutions to $\delta\Psi=0$. Let
us turn on $X^I,\ \ I=1,2,3, 4$, while setting $X^I,\ I=5,6,7,8$, to
zero and denote non-zero $X$'s as $X^i$, $i=1,2,3,4$. The BPS
equation takes the form%
\be\label{1/4BPS}%
\left(\gamma^\mu \Gamma^i D_\mu
X^i+\frac{1}{6}\Gamma^{ijk}[X^i,X^j,X^k, T]\right)\epsilon=0\ .
\ee%
The above is basically the Basu-Harvey equation \cite{BH}. Here we
just review its solutions. Consider the configurations for which the
gauge fields are vanishing and also take $X^i$ to only depend on one
of worldvolume coordinates, say $x_2$. The $x$ dependence of the two
terms in \eqref{1/4BPS} can be factored out if and only if
\footnote{Note that in our conventions the scalar fields $X^I$ have mass dimension
$1/2$, while fermions $\Psi$ and gauge fields $A_{\mu AB}$  have mass dimension $1$.}%
\be\label{1/4BPS-x-dependence}%
X^i=\frac{1}{\sqrt{2\cdot |x_2-x_2^0|}} J^i ,%
\ee%
where $x_2^0$ is an integration constant and $J^i$ are
some ($x$-independent) matrices which should satisfy%
\be\label{J-eq}%
\left(\gamma^2 \Gamma^i J^i-\frac{s}{6}[J^i,J^j, J^k,
T]\Gamma^{ijk}\right)\epsilon=0\ , %
\ee%
where $s=\frac{x_2-x_2^0}{|x_2-x_2^0|}$ is taking $\pm 1$ values.
The $\Gamma^i$ are four of the $SO(8)$ Majorana-Weyl Dirac
matrices and hence can be viewed as $SO(4)\in SO(8)$ Dirac
matrices and therefore%
\[
\Gamma^{ijk}=\epsilon^{ijkl} \Gamma^5\Gamma^l ,
\]
where $\Gamma^5$ is the $SO(4)$ chirality matrix. $\epsilon$ is a
two component $3d$ fermion, while also in $\mathbf{8}_c$ of $SO(8)$
R-symmetry. As such,%
\be\label{3d-4d-chirality}%
\begin{split}%
\gamma^2\epsilon &=s_1\epsilon\ ,\cr \Gamma^5\epsilon &=s_2
\epsilon\ ,
\end{split}%
\ee%
where $s_1$ and $s_2$ can (independently) be $+1$ or $-1$.
Inserting
the above into \eqref{J-eq} and after some simple algebra we arrive at%
\be\label{J-simplified}%
[J^i, J^j, J^k, T]= s s_1 s_2 \epsilon^{ijkl}J^l\ .
\ee%
The above has a solution in terms of $2N\times 2N$ representation of
$SO(4)$, if we take $J^i$ to be proportional to $2N\times 2N$
$SO(4)$ Dirac matrices and $T$ to be proportional to $2N\times 2N$
``$\Gamma^5$''. (For a detailed discussion on constructing solutions
of \eqref{J-simplified} see \cite{TGMT}.) Moreover, for $s=+1$ (i.e.
for $x_2>x_2^0$) we should take $s_1s_2=-1$ and for $s=-1$
$s_1s_2=+1$. Let us focus on the $s=+1$ for which there are two
types of solutions, $s_1=+1,\ s_2=-1$, or $s_1=-1,\ s_2=+1$, each of
which are invariant under transformations generated by four
independent $\epsilon$'s and hence altogether our solution is
invariant under eight fermionic transformations. (For the $s=-1$,
$x_2<x_2^0$ case, there are again eight $\epsilon$'s.)

We should now check if our configurations indeed satisfy the closure
of the two successive \susy\ transformations. To see this we note
that, $v^\mu\partial_\mu X^I=-2i\bar\epsilon_2 \gamma^2\epsilon_1
\partial_2 X^I$. On the other hand for our solutions $\epsilon_i$
are eigenstates of $\gamma^2$ (\emph{cf.} \eqref{3d-4d-chirality})
and therefore, $v^\mu|_{\mu=2}=0$. For the same reason $V_{JK}$
\eqref{vmu-VJK} is zero and hence $[\delta_1,\delta_2]
X^I=v^\mu\partial_\mu X^I=0$. As a result our configuration is a
1/4 BPS configuration and preserves 8 supercharges.

It is instructive to also present the solution in terms of our
earlier notation and $(X^i)^\pm$ components: $(X^i)^\pm \propto
(1\pm T) {\cal J}^i$, where $T={\cal J}^5$ and ${\cal J}$'s are
$2N\times 2N$  $SO(4)$ Dirac $\gamma$-matrices \cite{TGMT}. It is
evident that $((X^i)^+)^\dagger=(X^i)^-$ and moreover for our
solution $X^i_+=X^i_-$. In terms of the ABJM complex $Z_\alpha$
fields our solution is $Z_\alpha={\bar Z}_{\alpha}=X^i$. Note also
that our solution is invariant under parity.

\section{Discussion}

In this work we have attempted generalizing the \3dNe BLG gauge
theory by extending the notion of three-algebras. As we argued
invariance of the BLG action under gauge symmetry requires a weaker
condition than what is demanded by Bagger-Lambert (BL)
three-algebras. In particular, in this work we focused on a notion
of \emph{extended \FI }.  Based on this notion we constructed an
extended three-algebra, while giving a representation of the BL
three-brackets in terms of an explicitly totally antisymmetric
four-bracket and an explicit matrix representation for the algebra
elements.

We showed that the closure of our extended three-algebra, under the
working assumption that $t^A$ \eqref{TApm-T} are generators of a
(semi-simple) Lie-algebra, fixes $t^A$ to be generators of $u(N)$ in
its $N\times N$ (fundamental) representation. We hence called this
new three-algebra, the $u(N)$-based extended three-algebra. As we
showed (see appendix C) the $N=2$ case  reproduces two copies of the
BL $so(4)$-based three-algebra and in this sense our extended
algebras are a generalization of BL three-algebras to $N>2$ (in the
M2-brane picture $N$ is the number of M2-branes). It is interesting
to explore whether one can relax this working assumption and study
other kinds of extended three-algebras which may arise in this way
and the BLG theory based on them.

We showed that the BLG theory for the $u(N)$-based extended
three-algebra\  can be rewritten in terms of a $3d$ \sun2 \cs theory
with $SO(8)$ global symmetry and fields in the bi-fundamentals of
the \sun2. Our theory, however, has  twice more than the expected
physical degrees of freedom. The bi-fundamental fields appear as a
direct result of our choice of $2N\times 2N$ matrices (\emph{cf.}
footnote 12).

 This theory, although invariant under the $3d$ parity,
involves propagating scalar fields which are not parity invariant.
To reduces the number of scalar degrees of freedom to the desired
one, half of the existing ones, and also to close the fermionic
variations onto a \susy\ algebra, we projected the states onto the
$SU(4)\times U(1)\in SO(8)$ sector of the Hilbert space which is
invariant under the parity times the $U(1)_\xi$ charge conjugation.
After this projection the theory becomes an \Ns \sun2\ \cs\ theory.
We discussed connection of our model with that of ABJM \cite{ABJM}.
As discussed, for the special $N=2$ case there is another way of
projecting out half of the extra degrees of freedom in an $SO(8)$
invariant manner and obtain the original Bagger-Lambert theory. It
is interesting to see if there are other ways of projecting the
extra degrees of freedom by the other discrete symmetries of our
problem and obtain other $3d$ supersymmetric possibly $SO(8)$
invariant \cs theories.

Although the ${\cal N}=6$ \cs theory is very restrictive
\cite{N=6-classification}, there are other possibilities (than
\sun2) for the gauge groups and matter content. Moreover, motivated
by the ABJM model, recently many supersymmetric \cs theories with
${\cal N}\leq 5$ has been constructed (e.g. see \cite{Less-SUSY} and
references therein). As we showed, for the ${\cal N}=6$ theories,
the cases others than \sun2 theory does not have a representation in
terms of our extended three-algebras. In this viewpoint the ABJM
type theory is special. It is interesting to see whether within our
extended algebras (presumably by relaxing the working assumption
mentioned above) or within the ``generalized Bagger-Lambert
three-algebras'' \cite{BL3-bracket-N6} these other cases also find a
representation in terms of three-algebras. For a recent work in this
direction see \cite{Yamazaki:2008gg}.



We gave a very suggestive picture for realization of \sun2 gauge
group, our argument was a generalization or extension of the similar
picture for D-branes. It is desirable to make our ``pair-wise''
picture more quantitative and see how the structure of the extended
three-algebra may come out of this picture.

To provide further evidence one may also construct other BPS
configurations and compare it against the result expected from a
system of M2-branes. One may also compute the supersymmetric
(Witten) indices for our \sun2 theory. The computation should
closely follow that of the ABJM theory \cite{Shiraz}. However, in
our case we should also implement the ``projection onto
supersymmetric Hilbert space'' in computation of the partition
function or supersymmetric indices. Providing these further pieces
of evidence in support of our proposed model is postponed to
future works.%
\vskip 5mm

{\large{\bf Acknowledgements}}

The author is indebted to Neil Lambert for his clarifying comments
and critiques about the \susy\ closure. I would like to thank Ofer
Aharony, Mohammad Ali-Akbari and  Joan Simon  for fruitful comments
or discussions.


\appendix

\section{Conventions and useful identities for $su(N)$ algebras}

In our conventions, generators of the $u(N)$ algebra in its
$N\times N$ (fundamental) representation are denoted by $t^A$,
$A=0,1,\cdots, N^2-1$. Among $t^A$'s, $t^0$ is the generator of
$u(1)$ and $t^a$, $a=1,2,\cdots, N^2-1$, are generators of
$su(N)$. In our normalization%
\be\label{tA-normalize}%
Tr(t^A t^B)=\half \delta^{AB},%
\ee%
and therefore%
\be%
t^0=\frac{1}{\sqrt{2N}} \id_N. %
\ee%
\textbf{The product of two generators:}%
\be\label{tAB-prodcut}%
\begin{split}%
t^a t^b &=\frac{i}{2} f^{abc}\ t^c + \half d^{abc} t^c+
\frac{1}{2N} \delta^{ab}\id_N\ ,\\ t^0 t^a &=t^a
t^0=\frac{1}{\sqrt{2N}} t^a,\\
t^0 t^0&=\frac{1}{2N} \id\ ,
\end{split}\ee%
where $f^{abc}$ (which is totally anti-symmetric) is the structure
constant of the $su(N)$ algebra and $d^{abc}$ is the totally
symmetric traceless tensor of $su(N)$. From the above it is seen
that%
\[
\sum_A\ t^A t^A=\frac{N}{2} \id_N\ ,
\]
and
\[
[t^a, t^b]=if^{abc} t^c,\qquad \{t^a, t^b\}=d^{abc} t^c+
\frac{1}{N} \delta^{ab} \id\ .\] {\textbf{Useful identities on the
product of $f$'s and $d$'s:}}

Here we list some  identities which have been used in computations
performed in the main text. These identities are taken from
\cite{McFarlane}.

\begin{itemize}
\item Product of two $f$'s or $d$'s:
\be\label{two-f-d}%
\begin{split}
f_{acd}\ f_{bcd} &= N\ \delta_{ab}\ ,\\
f_{acd}\ d_{bcd} &= 0\ ,\\
d_{acd}\ d_{bcd} &= \frac{N^2-4}{N}\ \delta_{ab}\ .
\end{split}%
\ee%
\item {The Jacobi identities}
\be\label{Jacobi}%
\begin{split}
f_{ade} f_{bce}+f_{bde} f_{cae}+ f_{cde} f_{abe} &= 0 ,\\
f_{ade} d_{bce}+f_{bde} d_{cae}+ f_{cde} d_{abe} &= 0 ,\\
f_{abe} f_{cde} =
\frac{2}{N}(\delta_{ac}\delta_{bd}-\delta_{ad}\delta_{bc})+&
(d_{ace}d_{bde}-d_{ade}d_{bce})\ .
\end{split}%
\ee%
\item Product of three $f$'s or $d$'s
\be\label{three-f-d}%
\begin{split}
f_{ade}f_{beg}f_{cgd}&=\frac{N}{2} f_{abc}\ ,\\
d_{ade}f_{beg}f_{cgd}&=\frac{N}{2} d_{abc}\ ,\\
d_{ade}d_{beg}f_{cgd}&=-\left(\frac{N^2-4}{2N}\right) f_{abc}\ ,\\
d_{ade}d_{beg}d_{cgd}&=3\left(\frac{N^4-4}{2N}\right) d_{abc}\ .
\end{split}%
\ee%
(For the last identity there is a typo in \cite{McFarlane} which
we have corrected.)

\end{itemize}

\section{On the uniqueness of the $u(N)$-based extended
three-algebras}

Here we present line of arguments which show that with the working
assumption that $t^A$ are generators of semi-simple Lie-algebras,
(3.11) can only hold for $u(N)$ in its $N\times N$ representation.
Our argument is arranged in two steps:

 { \textbf{I})} For any finite dimensional matrix representation of
simple Lie-algebra the generators are traceless (because trace of
a commutator is zero). On the other hand one can always normalize
the basis such that $Tr(t^A t^B)=\frac12\delta_{AB}$, $t^A$ being
generators of any simple algebra. Trace of left-hand-side of
(3.11) is not zero (it is just the structure constant of the
algebra $f^{ABC}$). Therefore, $t^A$ satisfying (3.11) cannot be
generators of any simple \emph{non-Abelian} Lie-algebra or direct
products of thereof. Moreover, to satisfy (3.11) for a
``semi-simple'' Lie algebra generators it must contain
\emph{Abelian} factors.

\textbf{II}) One can show that in order (3.11) to hold,
generically, the product of any two generators, and not only their
commutators,
should also be in the same algebra, i.e. %
\be\label{tAtB}%
\{t^A, t^B\}=F^{ABC}t^C,\ %
\ee%
for some numeric coefficient expansions $F^{ABC}$. In the
matrix representations, this latter only holds only for any
generic $N\times N$ matrices and within our working assumption
that is only $u(N)$ (or direct products of $u(N)$'s).

To see how (3.11) leads to \eqref{tAtB}, let us assume that we are
working with $N\times N$ representation for $t^A$'s and %
\be%
\{t^A, t^B\}=F^{ABC}t^C+G^{AB\alpha} X^{\alpha},%
\ee%
where $X^{\alpha}$ are the set of all $N\times N$ matrices which
cannot be expressed as linear combination of $t^A$'s. In other
words, $X^\alpha$ are ``complementary'' to $t^A$ in covering the
$N\times N$ matrices. Without loss of generality we may choose the
$X^{\alpha}$ such that $Tr(t^A X^\alpha)=0$, and let $Tr(X^\alpha
X^\beta)=g^{\alpha\beta}$. Next, multiply both sides of (3.11) by
$X^\alpha$ and take the trace. The right-hand-side vanishes while
the left-hand-side does not; it vanishes only if $G^{AB\alpha}=0$
(for any $A,B, \alpha$) or $g^{\alpha\beta}=0$ (for any
$\alpha,\beta$). The latter cannot happen because there is a
simple counter-example: if the $t^A$ are not generators of $u(N)$,
then there are elements in the ``complementary'' set the trace of
product of its generators are not zero. We then remain with
$G^{AB\alpha}=0$ choice which implies \eqref{tAtB} and hence
proving the statement.

\section{$so(4)$-based Bagger-Lambert three-algebra as an extended three-algebra}

As mentioned, in our construction the $u(N)$-based extended
three-algebra is a metric three-algebra with a positive definite
metric. This is readily seen from \eqref{TA-normalization}. (For the
same reason we do not expect the Lorentzian $u(N)$ three-algebras to
have a realization in terms of our extended three-algebras.
Nonetheless, as discussed in \cite{ASS}, they do admit a
representation in terms of matrices and four-brackets.) It is hence
interesting to see if the $so(4)$-based Bagger-Lambert (BL)
three-algebra can be obtained as a special case of our $u(N)$-based
extended three-algebra.

The obvious candidate for realization of $so(4)$-based BL
three-algebra is $u(2)$-based extended three-algebra. For this
case the
$T^A_{\pm}$ generators are%
\be\label{u2-generators}%
T^A_\pm=\half \sigma^A \otimes \sigma_\pm\ ,  \qquad A=0,1,2,3.%
\ee%
where $\sigma^A=( \id_2, \sigma^a),\ a=1,2,3$.
The above are eight matrices and can be decomposed as%
\be\label{gamma-mu-u2}%
\begin{split}%
T^a_\pm &=\frac14 \gamma^a (1\pm \gamma^5),\ \qquad a=1,2,3\ , \cr %
T^0_\pm &=\pm \frac{1}{4i}\gamma^4 (1\mp \gamma^5),\
\end{split}%
\ee%
where $T=1\otimes \sigma^3=\gamma^5$. $(T^a_++T^a_-)$ and
$i(T^0_+-T^0_-)$ combination of the $T^A_\pm$ matrices, are
 the $so(4)$ Dirac $\gamma$-matrices. In other words, if we restrict
 ourselves to the sector of the theory in which $\Phi^a_+=\Phi^a_-$
 and $\Phi^0_+=-\Phi^0_-$, the $u(2)$-based extended three-algebra becomes the $so(4)$-based BL
three-algebra written in another basis. There are, however, some
comments:

1) The $su(2)$ algebra, among the $su(N)$ algebras, is special in
the sense that its totally symmetric traceless three tensor
$d_{abc}$ identically vanishes (which is compatible with
\eqref{two-f-d} and \eqref{three-f-d} identities). This brings about
a great simplification in the structure constants $f^{ABCD}$.

2) As we can see among eight $T^A_\pm$ one can construct
$\gamma^\mu$ and $\gamma^\mu\gamma^5$ ($\mu=1,2,3,4)$ and one can
restrict the elements of the algebra to have components along
$\gamma^\mu$ or along $\gamma^\mu\gamma^5$. In this sense our
$u(2)$-based extended three-algebra contains two copies of the
$so(4)$-based Bagger-Lambert three-algebra. One can choose to work
with one half, say the one spanned by $\gamma^\mu$'s, as they close
onto a sub-three-algebra. In this subalgebra, the structure
constants take the form $\epsilon^{\mu\nu\alpha\beta}$. For the same
reason in the $u(2)$ case in this specific sector our ``extended \FI
'' becomes the standard \FI .


\section{Compatibility of supersymmetry and parity}%

Fermionic transformations \eqref{SUSY-trans-four-bracket} are
compatible with parity  if the following identities are
satisfied%
\bse\label{parity-susy-compatible}%
\begin{align}%
(\delta_{susy} X^I)_{parity}&=\delta_{susy} (X^I_{parity})\ ,\\
(\delta_{susy} \Psi)_{parity}&=\delta_{susy} (\Psi_{parity})\ ,\\
(\delta_{susy} {\tilde A}_{\mu AB})_{parity}&=\delta_{susy}
({\tilde A}_{\mu AB})^*\ ,
\end{align}%
\ese%
where in the last equality we have used
\eqref{parity-gauge-field-cc}, $X^I_{parity}$, $\Psi_{parity}$ are
defined in \eqref{XI-Psi-Parity} and note that under parity the
\susy \ parameter $\epsilon_p$ is transformed as \footnote{We
would like to comment that the \susy\ transformations of the
Bagger-Lambert theory \cite{BL2} are compatible with parity in the
sense of \eqref{parity-susy-compatible} with the same choice for
$\epsilon_p$.}%
\be\label{epsilon-parity}%
\epsilon\longrightarrow \epsilon_p=-\gamma^2\epsilon\ .%
\ee%
With this choice $\bar\epsilon\ \Psi$ behaves as a scalar (rather
than a pseudoscalar) and $\bar\epsilon\gamma^\mu \Psi$ behaves as
a vector. Therefore, recalling (\ref{SUSY-trans-four-bracket}a)
and \eqref{XI-Psi-Parity}, (\ref{parity-susy-compatible}a) becomes
immediate.

To check (\ref{parity-susy-compatible}b), we note that
\[%
\gamma^\mu(D_\mu X^I)_{parity}= -\gamma^2\gamma^\mu D_\mu
(X^I_{parity})\gamma^2\ ,%
\]%
and hence the first term in $(\delta_{susy} \Psi)$, goes to the
first term in $\delta_{susy} (\Psi_{parity})$. Recalling
\eqref{bracket-parity} one finds that the second term in
$\delta_{susy}\Psi$ goes to
$\delta_{susy} (\Psi_{parity})$. Putting these together we have:%
\[
(\delta_{susy}\Psi)_{parity}=\gamma^2\left(\gamma^\mu D_\mu
X^I_{parity}\Gamma^I\epsilon-\frac{1}{6}[X^I_{parity},X^J_{parity},X^K_{parity}, T]\Gamma^{IJK}\epsilon\right) .%
\]%
which is nothing but (\ref{parity-susy-compatible}b).

To verify (\ref{parity-susy-compatible}c) we note that%
\be\label{gauge-field-susy}%
\delta_{susy} {\tilde A}_{\mu CD}=i\bar\epsilon \gamma_\mu
\Gamma^I \biggl((X^I)^+_A\ \Psi^-_B-(X^I)^-_B\ \Psi^+_A\biggr)
f_{ABCD}\ , %
\ee%
and hence%
\be%
\begin{split}%
(\delta_{susy} {\tilde A}_{\mu CD})_{parity} &=i\bar\epsilon
\gamma^2\gamma_\mu\gamma^2 \Gamma^I \biggl((X^I)^-_A\
\Psi^+_B-(X^I)^+_B\ \Psi^-_A\biggr)
f_{ABCD}\ \cr %
&=-i\bar\epsilon \gamma^2\gamma_\mu\gamma^2 \Gamma^I
\biggl((X^I)^+_A\ \Psi^-_B-(X^I)^-_A\ \Psi^+_B\biggr)
f_{ABDC}\ \cr %
&= \delta_{susy} ({\tilde A}_{\mu CD})_{parity}\ ,
\end{split}%
\ee%
where in the second line of the above we have used
\eqref{f-symmetries} and in the third line
\eqref{parity-gauge-field}. Note also that
$\gamma^2\gamma_\mu\gamma^2\equiv -\gamma^p_\mu$ where
$\gamma^p_\mu$ is equal to $\gamma^0,\gamma^1$ for $\mu=0,1$ and
to $-\gamma^2$ for $\mu=2$.



\begin{thebibliography}{99}

\bibitem{Schwarz-04}
  J.~H.~Schwarz,
  ``Superconformal Chern-Simons theories,''
  JHEP {\bf 0411}, 078 (2004)
  [arXiv:hep-th/0411077].

\bibitem{BL1}
  J.~Bagger and N.~Lambert,
  ``Modeling multiple M2's,''
  Phys.\ Rev.\  D {\bf 75} (2007) 045020
  [arXiv:hep-th/0611108].

\bibitem{BL2}
  J.~Bagger and N.~Lambert,
  ``Gauge Symmetry and Supersymmetry of Multiple M2-Branes,''
  Phys.\ Rev.\  D {\bf 77} (2008) 065008
  [arXiv:0711.0955 [hep-th]].

J.~Bagger and N.~Lambert,
  ``Comments On Multiple M2-branes,''
  JHEP {\bf 0802}, 105 (2008)
  [arXiv:0712.3738 [hep-th]].


 \bibitem{Gust1}
  A.~Gustavsson,
  ``Algebraic structures on parallel M2-branes,''
  arXiv:0709.1260 [hep-th].


\bibitem{Gust2}
  A.~Gustavsson,
  ``Selfdual strings and loop space Nahm equations,''
  arXiv:0802.3456 [hep-th].

\bibitem{Lambert-strings08}%
N. Lambert,`` Lagrangians for Multiple M2-branes,'' \emph{the talk
presented in Strings 2008}, CERN August 2008,
\texttt{http://www.cern.ch/Strings2008}.


\bibitem{Papadopoulos:2008sk}
J.~P.~Gauntlett and J.~B.~Gutowski,
  ``Constraining Maximally Supersymmetric Membrane Actions,''
  arXiv:0804.3078 [hep-th].

 G.~Papadopoulos,
  ``M2-branes, 3-Lie Algebras and Plucker relations,''
  arXiv:0804.2662 [hep-th].



\bibitem{Russo}
  J.~Gomis, G.~Milanesi and J.~G.~Russo,
  ``Bagger-Lambert Theory for General Lie Algebras,''
  arXiv:0805.1012 [hep-th].


\bibitem{Verlinde}
  S.~Benvenuti, D.~Rodriguez-Gomez, E.~Tonni and H.~Verlinde,
  ``N=8 superconformal gauge theories and M2 branes,''
  arXiv:0805.1087 [hep-th].

\bibitem{HIM}
P.~M.~Ho, Y.~Imamura and Y.~Matsuo,
  ``M2 to D2 revisited,''
  JHEP {\bf 0807}, 003 (2008)
  [arXiv:0805.1202 [hep-th]].


\bibitem{Jose-Fig}
J.~Figueroa-O'Farrill, P.~de Medeiros and E.~Mendez-Escobar,
  ``Lorentzian Lie 3-algebras and their Bagger-Lambert moduli space,''
  arXiv:0805.4363 [hep-th];
  ``Metric Lie 3-algebras in Bagger-Lambert theory,''
  arXiv:0806.3242 [hep-th].








\bibitem{Sen1}
  S.~Cecotti and A.~Sen,
  ``Coulomb Branch of the Lorentzian Three Algebra Theory,''
  arXiv:0806.1990 [hep-th].

\bibitem{Sen2}
  S.~Banerjee and A.~Sen,
  ``Interpreting the M2-brane Action,''
  arXiv:0805.3930 [hep-th].

\bibitem{ghost-gauging}
M.~A.~Bandres, A.~E.~Lipstein and J.~H.~Schwarz,
  ``Ghost-Free Superconformal Action for Multiple M2-Branes,''
  arXiv:0806.0054 [hep-th].

J.~Gomis, D.~Rodriguez-Gomez, M.~Van Raamsdonk and H.~Verlinde,
  ``The Superconformal Gauge Theory on M2-Branes,''
  arXiv:0806.0738 [hep-th].

 B.~Ezhuthachan, S.~Mukhi and C.~Papageorgakis,
  ``D2 to D2,''
  arXiv:0806.1639 [hep-th].

\bibitem{ASS}%
M.~Ali-Akbari, M.~M.~Sheikh-Jabbari and J.~Simon,
  ``Relaxed Three-Algebras: Their Matrix Representations and Implications for
  Multi M2-brane Theory,''
  arXiv:0807.1570 [hep-th].


\bibitem{VerlindeM2scattering}
H.~Verlinde,
  ``D2 or M2? A Note on Membrane Scattering,''
  arXiv:0807.2121 [hep-th].

\bibitem{AdS/CFT}
  J.~M.~Maldacena,
  ``The large N limit of superconformal field theories and supergravity,''
  Adv.\ Theor.\ Math.\ Phys.\  {\bf 2}, 231 (1998)
  [Int.\ J.\ Theor.\ Phys.\  {\bf 38}, 1113 (1999)]
  [arXiv:hep-th/9711200].

 \bibitem{Mark}
  M.~Van Raamsdonk,
  ``Comments on the Bagger-Lambert theory and multiple M2-branes,''
JHEP {\bf 0805}, 105 (2008),  arXiv:0803.3803 [hep-th].




\bibitem{ABJM}
  O.~Aharony, O.~Bergman, D.~L.~Jafferis and J.~Maldacena,
  ``N=6 superconformal Chern-Simons-matter theories, M2-branes and their
  gravity duals,''
  arXiv:0806.1218 [hep-th].

\bibitem{BL3-bracket-N6}
  J.~Bagger and N.~Lambert,
  ``Three-Algebras and N=6 Chern-Simons Gauge Theories,''
  arXiv:0807.0163 [hep-th].


\bibitem{TGMT}
  M.~M.~Sheikh-Jabbari,
  ``Tiny graviton matrix theory: DLCQ of IIB plane-wave string theory, a
  conjecture,''
  JHEP {\bf 0409}, 017 (2004)
  [arXiv:hep-th/0406214].


  M.~M.~Sheikh-Jabbari and M.~Torabian,
  ``Classification of all 1/2 BPS solutions of the tiny graviton matrix
  theory,''
  JHEP {\bf 0504}, 001 (2005)
  [arXiv:hep-th/0501001].




\bibitem{Takhtajan}
  L.~Takhtajan,
  ``On Foundation Of The Generalized Nambu Mechanics (Second Version),''
  Commun.\ Math.\ Phys.\  {\bf 160}, 295 (1994)
  [arXiv:hep-th/9301111].


\bibitem{Cherkis}
  S.~Cherkis and C.~Saemann,
  ``Multiple M2-branes and Generalized 3-Lie algebras,''
  arXiv:0807.0808 [hep-th].


\bibitem{JoseF-O}
 P.~de Medeiros, J.~Figueroa-O'Farrill, E.~Méndez-Escobar,
 P.~Ritter, `` On the Lie-algebraic origin of metric 3-algebras,''
 arXiv:0809.10860 [hep-th].

\bibitem{Witten}
  E.~Witten,
  ``Bound states of strings and p-branes,''
  Nucl.\ Phys.\  B {\bf 460}, 335 (1996)
  [arXiv:hep-th/9510135].


  \bibitem{M2onOrbifold}
  N.~Lambert and D.~Tong,
  ``Membranes on an Orbifold,''
  arXiv:0804.1114 [hep-th].

  J.~Distler, S.~Mukhi, C.~Papageorgakis and M.~Van Raamsdonk,
  ``M2-branes on M-folds,''
  arXiv:0804.1256 [hep-th].





\bibitem{BH}

 A.~Basu and J.~A.~Harvey,
  ``The M2-M5 brane system and a generalized Nahm's equation,''
  Nucl.\ Phys.\  B {\bf 713}, 136 (2005)
  [arXiv:hep-th/0412310].

\bibitem{Terashima}
  S.~Terashima,
  ``On M5-branes in N=6 Membrane Action,''
  JHEP {\bf 0808}, 080 (2008)
  [arXiv:0807.0197 [hep-th]].

K.~Hanaki and H.~Lin,
  ``M2-M5 Systems in N=6 Chern-Simons Theory,''
  JHEP {\bf 0809}, 067 (2008)
  [arXiv:0807.2074 [hep-th]].

\bibitem{Other-solutions}
 C.~Krishnan and C.~Maccaferri,
  ``Membranes on Calibrations,''
  JHEP {\bf 0807}, 005 (2008)
  [arXiv:0805.3125 [hep-th]].

 T.~Fujimori, K.~Iwasaki, Y.~Kobayashi and S.~Sasaki,
  ``Time-dependent and Non-BPS Solutions in N=6 Superconformal Chern-Simons
  Theory,''
  arXiv:0809.4778 [hep-th].





\bibitem{N=6-classification}%
K.~Hosomichi, K.~M.~Lee, S.~Lee, S.~Lee and J.~Park,
  ``N=5,6 Superconformal Chern-Simons Theories and M2-branes on Orbifolds,''
  arXiv:0806.4977 [hep-th].

 M.~Schnabl and Y.~Tachikawa,
  ``Classification of N=6 superconformal theories of ABJM type,''
  arXiv:0807.1102 [hep-th].


\bibitem{Less-SUSY}

K.~Hosomichi, K.~M.~Lee, S.~Lee, S.~Lee and J.~Park,
  ``N=4 Superconformal Chern-Simons Theories with Hyper and Twisted Hyper
  Multiplets,''
  JHEP {\bf 0807}, 091 (2008)
  [arXiv:0805.3662 [hep-th]].

O.~Aharony, O.~Bergman and D.~L.~Jafferis,
  ``Fractional M2-branes,''
  arXiv:0807.4924 [hep-th].

\bibitem{Yamazaki:2008gg}
  M.~Yamazaki,
  ``Octonions, $G_2$ and generalized Lie 3-algebras,''
  arXiv:0809.1650 [hep-th].


\bibitem{schwarz}
  M.~A.~Bandres, A.~E.~Lipstein and J.~H.~Schwarz,
  ``N = 8 Superconformal Chern--Simons Theories,''
  JHEP {\bf 0805}, 025 (2008)
  [arXiv:0803.3242 [hep-th]].

\bibitem{Bandres:2008ry}
  M.~A.~Bandres, A.~E.~Lipstein and J.~H.~Schwarz,
  ``Studies of the ABJM Theory in a Formulation with Manifest SU(4)
  R-Symmetry,''
  JHEP {\bf 0809}, 027 (2008)
  [arXiv:0807.0880 [hep-th]].


\bibitem{Shiraz}
  J.~Bhattacharya and S.~Minwalla,
  ``Superconformal Indices for ${\cal N}=6$ Chern Simons Theories,''
  arXiv:0806.3251 [hep-th].

\bibitem{McFarlane}%
A.J.~ MacFarlane, A.~Sudbery and P.H.~Weisz, ``On Gell-Mann's
$\lambda$-matrices, $d$ and $f$ tensors, Octets, and
parametrizations of $su(3)$,'' Com. Math. Phys. {\bf 11} (1968)
77.



\end{thebibliography}
\end{document}